\documentclass[aps,prb,reprint,groupedaddress,showpacs]{revtex4-1}

\usepackage{dcolumn}
\usepackage{graphicx}

\usepackage{color}
\usepackage[version=3]{mhchem}

\usepackage{subcaption}

\usepackage{multirow}

\begin{document}
\newcommand{\vect}[1]{\boldsymbol{#1}}
\newcommand{\kpoint}{$\vect{k}$-point}
\newcommand{\kpoints}{$\vect{k}$-points}


\title{Oxidation of GaN: An ab initio thermodynamic approach}


\author{Adam J. Jackson}
\author{Aron Walsh}
\email[]{a.walsh@bath.ac.uk}
\affiliation{Centre for Sustainable Chemical Technologies \& Department of Chemistry,
University of Bath BA2 7AY,
United Kingdom}


\date{\today}

\begin{abstract} 
GaN is a wide-bandgap semiconductor used in high-efficiency LEDs and solar
cells. The solid is produced industrially at high chemical purities by deposition 
from a vapour phase, and oxygen may be included at this stage. Oxidation represents 
a potential path for tuning its properties without introducing more exotic 
elements or extreme processing conditions. 
In this work, \emph{ab initio} computational methods are used to examine the energy 
potentials and electronic properties of different extents of oxidation in GaN. 
Solid-state vibrational properties of Ga, GaN, \ce{Ga2O3} and a single
substitutional oxygen defect have been studied using the harmonic
approximation with supercells.
A thermodynamic model is outlined which combines the results of \emph{ab initio} 
calculations with data from experimental literature. 
This model allows free energies to be predicted for 
arbitrary reaction conditions within a wide process envelope.
It is shown that complete oxidation is favourable for all
industrially-relevant conditions, while the formation of defects can
be opposed by the use of high temperatures and a high \ce{N2}:\ce{O2}
ratio.
\end{abstract}


\maketitle
\section{\label{sec:intro}Introduction}
Solid-state lighting with light-emitting diodes (LEDs) offers exceptionally high efficiencies, and
systems with luminous efficacies of over 100 lmW$^{-1}$ are already 
commercially available.\cite{Murphy2012} 
With experimental systems achieving up to 169 lmW$^{-1}$, 
researchers are continuing to move performance towards the theoretical
 limit for white light of around 300 lmW$^{-1}$ 
 (depending on the definition of ``white'').\cite{Narukawa2010}
GaN, in pure and indium-doped forms, has been an instrumental part of this 
movement, forming many of the highest-performing LED systems.

Very pure semiconductors such as GaN are generally formed under high vacuum by
techniques including chemical vapour deposition (CVD) and molecular beam epitaxy
(MBE). Such conditions require specialised equipment and considerable energy. It would be desirable to carry out deposition reactions at more modest
pressures, but this risks the presence of gas impurities and may make the
system more difficult to control. In particular, oxygen is thought to form
solid solutions with GaN, substituting N atoms for O at low
concentrations ($<30$\%) and altering the resistivity and bandgap --
 important properties for its electronic applications.\cite{Aleksandrov2000}
500$^\circ$C is considered ``low-temperature'' for deposition and 800-1000$^\circ$C is more typical; this coincides with the maximum solubility of oxygen.\cite{Aleksandrov2000} 
A recent attempt at atomic layer deposition of GaN at modest temperatures ($<400^\circ$C) obtained a bulk oxygen concentration of 19.5\%.\cite{Ozgit2012}

However it has also been reported that higher temperatures can reduce the concentration of gallium oxide by controlling the rate of deposition; Obinata et al (2005) attribute the formation of Ga-O bonds to a film of ``excess Ga'', but also note that gallium oxide existed within their GaN films.\cite{Obinata2005} 
A proposed solution is annealing in the presence of ammonia, providing excess nitrogen.\cite{Sawada2007} 

In addition to the growth process, GaN is known to thermally decompose under vacuum at temperatures above around 700$^\circ$C (i.e. reaction conditions), with a strong temperature dependence, and this is also suppressed by nitrogen.\cite{Fernandez-Garrido2008}
The thermodynamic significance of varying nitrogen  pressures is therefore of interest.

A range of materials modelling techniques have been applied to GaN in
the past.
These include analytical pairwise potential\cite{Zapol1997, Catlow2010} and electronic structure studies\cite{Lany2010, Lambrecht1994, Limpijumnong2003, Carter2008}.
The computational defect physics was reviewed by Neugebauer
and Van de Walle\cite{Neugebauer1994}, while Zywietz \emph{et al.} investigated the incorporation
of oxygen on the material surface.\cite{Zywietz1999}
 
The key reaction considered in this study was the oxidation of GaN
under formation conditions. Complete oxidation is expected to occur at
high temperatures, above typical deposition conditions:
\[ 2\ce{GaN} + \tfrac{3}{2}\ce{O2} \rightarrow \ce{Ga2O3} + \ce{N2}
\]
while the dominant form of oxidation at deposition conditions is the
substitution of N atoms for O at low concentrations:
\[ \text{N}_{\text{N}} + \tfrac{1}{2}\ce{O2} \rightarrow \ce{O}_\ce{N} + \tfrac{1}{2}\ce{N2}
\]

\begin{figure}
\begin{subfigure}[b]{\columnwidth}
\includegraphics[width=4cm]{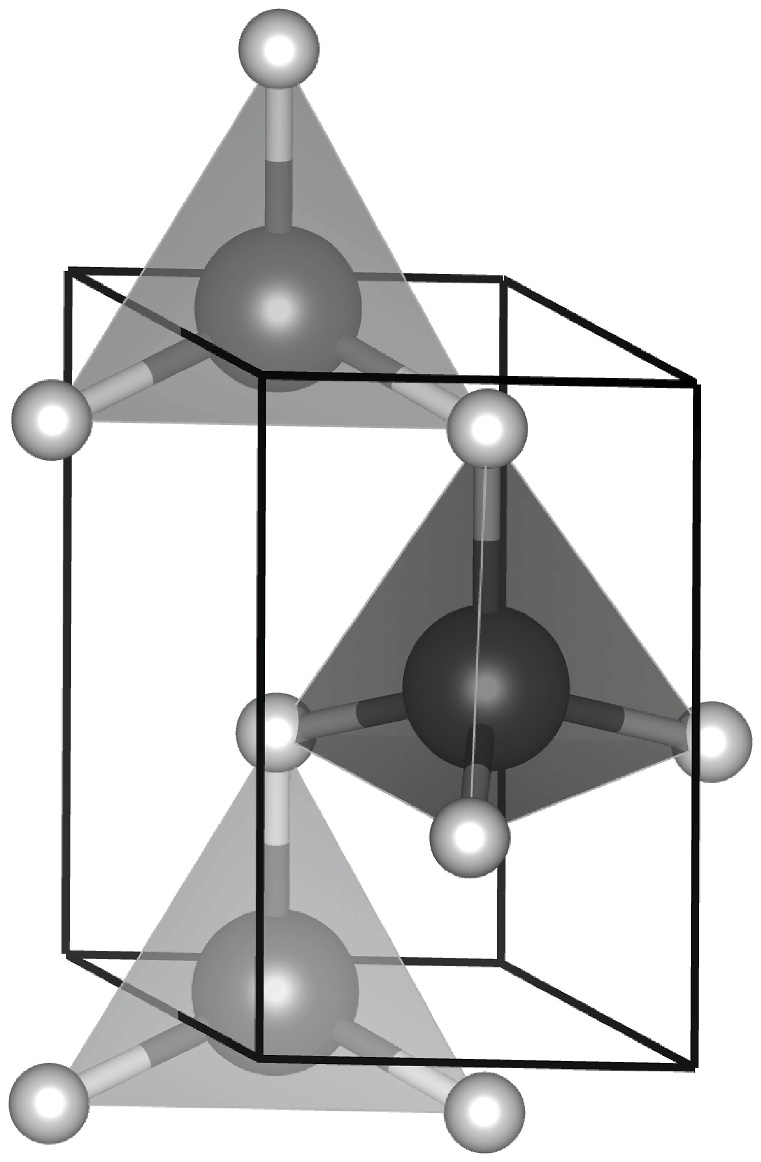}
\caption{}\label{fig:gan_cell}
\end{subfigure}

\begin{subfigure}[b]{\columnwidth}
\includegraphics[width=6cm]{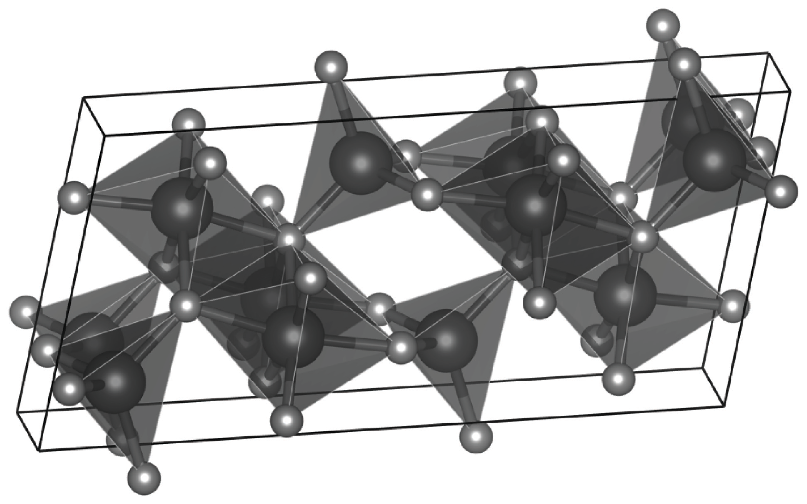}
\caption{}\label{fig:galox_cell}
\end{subfigure} 

\caption{Unit cell with bounding box for (a) GaN (b) $\beta$-\ce{Ga2O3}: dark spheres represent Ga atoms and lighter spheres show the position of N (a) and O (b) atoms.}
\end{figure}

GaN adopts the wurtzite crystal structure with tetrahedral ion coordination environments; this is represented by a small hexagonal unit
cell (Figure~\ref{fig:gan_cell}), 
whereas the stable $\beta$- structure of \ce{Ga2O3} corresponds to a more complex monoclinic unit cell,
with both tetrahedral and octahedral elements
(Figure~\ref{fig:galox_cell}).


\section{Methodology}

The aim of the study was to predict the envelope of conditions for
thermodynamically stable GaN, and free energies of several degrees of oxidation. This is achieved by using density functional theory (DFT)
to calculate the energies of pure and defective compounds.\cite{Hohenberg1964,Kohn1965} Energy minimisation
with DFT is based purely on the electronic potential field, and does not account
directly for any lattice vibrations, or the effect of pressure. By adding the
zero-point vibrational energy $E^{\text{ZP}}$ to the DFT-derived energy
$E^{\text{DFT}}$, we can obtain an energy value which we define as equivalent to various
thermodynamic potentials at zero temperature and zero pressure (indicated with a
superscript `0'):
\begin{equation}\label{eqn:atzero}
G^0 = H^0 = U^0 = E^0_{\text{potential}} + E^0_{\text{vibrational}}
    = E^{\text{DFT}} + E^{\text{ZP}}
\end{equation}
where $G$, $H$, $U$ are the Gibbs free energy, enthalpy and internal energy, respectively. The contributions of individual components may be considered in
terms of their chemical potentials $\mu_i$. For ideal materials the chemical
potential is equivalent to the Gibbs free energy of one unit (i.e. 1 mole) of the
pure material, and hence:
\begin{align}
\mu_i^0 = G_i^0 &= E_i^\textrm{DFT} + E_i^\textrm{ZP} \\
\intertext{Introducing the chemical potential at given reaction conditions $\mu_i(T,p)$, and rearranging:}
\mu_i(T,p) &= E_i^\textrm{DFT} + E_i^\textrm{ZP} + \left[\mu_i(T,p) - \mu_i^0\right] \label{eqn:potential_fromzero}
\end{align}

A slightly different approach is needed to collect this information for each
material. 
In the solid phase, $E_i^{\text{DFT}}$ is calculated for crystalline unit cells with a three-dimensional periodic boundary condition. $E_i^\textrm{ZP}$ and the free energy change with temperature and pressure 
$\left[\mu_i(T,p) - \mu_i^0\right]$ require some combination of approximations,
literature data and/or lattice dynamics calculations.\cite{Stoffel2010}

In the gas phase, the DFT energy for an isolated molecule must be found in a
method consistent with the solid component energies. Zero-point and free
energies are readily available in the literature for common gases such as
\ce{O2} and \ce{N2}, although care must be taken to use consistent reference points.

\subsection{Total energies and structures}
DFT calculations were carried out using the Fritz Haber
Institute \emph{ab initio} molecular simulations ({\sc FHI-aims})
package.\cite{Blum2009} {\sc FHI-aims} is highly scalable across thousands
of computer cores\cite{Havu2009}, and uses atom-centred
numerically-tabulated basis sets to describe all of the electrons in
the system. All energies and structures were converged using the
provided `tight' or `tier 2' basis set, which includes hydrogen-like s, p, d and f
atomic orbitals for Ga and adds a g orbital for N and O. 
With the exception of molecular oxygen, all calculations did not include spin-polarisation.
The PBEsol exchange-correlation functional was
selected; this functional uses the generalised gradient approximation
(GGA) and is intended for solid-state calculations.
\cite{Perdew2008,Csonka2009} The use of GGA for
examining the electronic structures of semiconductors has been
challenged recently as it tends to under-estimate formation energies and
bandgaps; nonetheless PBEsol is considered to offer a good balance of
accuracy and efficiency for total energies and structure
optimisation.\cite{Lany2008,Xiao2011}

\subsubsection{Pure compounds}
Inital crystal structures for GaN, $\beta$-\ce{Ga2O3} and Ga metal were obtained
from the literature via the Chemical Database Service at
Daresbury and the Inorganic Crystal Structure Database (ICSD).\cite{Paszkowicz2004,Ahman1996,* [{}] [{ [Note that due to an apparent error in the ICSD, the initial lattice parameters for Ga slightly deviate from this original source, which gives $a=4.5167$ \AA, $b=4.5107$ \AA, $c=7.6448$ \AA.]}] Bradley1935} The unit cell parameters and atomic
positions were converged with {\sc FHI-aims} and PBEsol to give energies
($E^\text{DFT}$) of the pure compounds. 
The geometry optimisation routine was permitted to vary both the cell contents and unit cell parameters in order to minimise the overall energy.
The routine employs analytical stress tensors with an adapted Broyden-Fletcher-Shanno-Goldfarb (BFGS) algorithm.\cite{Blum2009} 
The relaxed unit cell parameters are given in
Table~\ref{table:structures}.
\kpoints{} were defined as an evenly-spaced grid in reciprocal space, centred on
the $\Gamma$-point, and time-reversal symmetry was employed to reduce the
required number of calculations. The \kpoint{}-grid density was scaled to the unit cell
size to achieve uniform sampling with a target length cutoff of 10 \AA{}, as
described by Moreno and Soler.\cite{Moreno1992}

Oxygen (in the triplet spin configuration) and nitrogen gases were
modelled by setting an isolated pair of atoms 1~\AA{} apart and
allowing them to relax to a distance minimizing the energy. The
resulting distances are also included in Table~\ref{table:structures},
each overestimating their recorded spectroscopic value by 1\%
(Table~\ref{table:gas-zpe}).

\subsubsection{Defects}
Dilute oxidation in bulk GaN was modelled by the supercell approach: 72-atom,
128-atom and 300-atom supercells were created from the relaxed hexagonal 4-atom
GaN unit cell as described in Appendix~\ref{sec:transform}. Energies were calculated with
and without a single substitution of an N atom for an O atom.
The atomic positions within the cell were relaxed to find an energy minimum using the BFGS algorithm as above.

In order to model the dilute limit of oxidation, it is necessary here to apply a
band-filling correction. Oxygen substitution in GaN results in an
excess electron that occupies the conduction band.
As the defect concentration decreases, the conduction band filling
drops to the band minimum, usually at the gamma point. A correction
energy was calculated by integrating over the eigenvalues above this
reference energy for each \kpoint{}, following the method described by
Persson \emph{et al.}\cite{Persson2005} and discussed in more detail by Lany and Zunger.\cite{Lany2008a}
This has been implemented as a MATLAB routine, available on request.

\begin{table*}[tpb]  
\setlength{\extrarowheight}{2pt}
\caption{Relaxed structure parameters from DFT calculations with
  {\sc FHI-aims} and the PBEsol functional.
\kpoints{} are sampled evenly in reciprocal space and
  include the $\Gamma$-point. Lengths are in \AA{} and angles are in
  degrees. For diatomic gases $a$ is the distance between the nuclei.
\label{table:structures}}
\begin{ruledtabular}
\begin{tabular}{c d d d d d d  c d d d d d d}
 \multirow{2}{*}{Compound} &
 			\multicolumn{6}{c}{\text{Initial parameters}} & \multirow{2}{*}{\kpoints{}} & 
 	     	\multicolumn{6}{c}{\text{Relaxed structure}}  \\
  &			\multicolumn{1}{c}{$a$} & \multicolumn{1}{c}{$b$} &
  				 \multicolumn{1}{c}{$c$} & \multicolumn{1}{c}{$\alpha$} & 
  				 \multicolumn{1}{c}{$\beta$} & \multicolumn{1}{c}{$\gamma$} &  &
   		    \multicolumn{1}{c}{$a$} & \multicolumn{1}{c}{$b$} & 
   		    		\multicolumn{1}{c}{$c$} & \multicolumn{1}{c}{$\alpha$} & 
   		    		\multicolumn{1}{c}{$\beta$} & \multicolumn{1}{c}{$\gamma$} \\  \hline
GaN\cite{Paszkowicz2004} & 3.189 & 3.189 & 5.186 & 90 & 90 & 120 & 
[7 7 4] &  3.186 & 3.186 & 5.187 & 90.01 & 89.99 & 120.07 \\

\ce{Ga2O3}\cite{Ahman1996} & 12.214 & 3.037 & 5.798 & 90 & 103.83 & 90 &
 [2 8 4] & 	12.287 & 3.049 & 5.812 & 90.00 & 103.72 & 90.00 \\

Ga & 4.520 & 7.660 & 4.526 & 90 & 90 & 90 &
 [6 4 6] &  4.424 & 7.605 & 4.532 & 90.01 & 90.00 & 90.00 \\

\ce{O2} & 1.000 & & & & & &
	& 1.212 & & & & & \\

\ce{N2} & 1.000 & & & & & &
	& 1.101 & & & & & \\

\end{tabular}
\end{ruledtabular}
\end{table*}

\subsection{Gases - literature data}


\begin{table}
\setlength{\extrarowheight}{2pt}
\caption{Zero-point energies, standard enthalpies and bond lengths ($r$) for diatomic gases from literature\cite{Irikura2007,handbook-crc}
\label{table:gas-zpe}}
\begin{ruledtabular}
\begin{tabular}{c d d d d}
\multirow{2}{*}{Material} & \multicolumn{1}{c}{$E^{ZP}$} & 
		   \multicolumn{1}{c}{$E^{ZP}$} & 
\multicolumn{1}{c}{$\left[H^\theta - H^0 \right]$} &
	\multicolumn{1}{c}{$r$} \\
 & \multicolumn{1}{c}{/ eV} & \multicolumn{1}{c}{/ kJ mol$^{-1}$} &
 \multicolumn{1}{c}{/ kJ mol$^{-1}$} &
 \multicolumn{1}{c}{/ \AA} \\
\hline
\ce{O2}  & 0.0976       & 9.42    &  8.680  & 1.2075 \\
\ce{N2}  & 0.1458       & 14.07   &  8.670  & 1.0977 \\
\end{tabular}
\end{ruledtabular}
\end{table}

Gas properties in \emph{ab initio} thermodynamics can be calculated using statistical mechanics, but in practice are generally drawn from experimental values.\cite{Reuter2005,Reuter2001,Morgan2010}
In this case properties for \ce{O2} and \ce{N2} were calculated using data from standard thermochemical tables,
which have been fitted by NIST to polynomial equations of the form developed by Shomate. \cite{Chase1998,Linstrom2005,Shomate1944,Shomate1954} Such correlations are especially convenient for use in computer programs.
The correction for
temperature and pressure in Eq. (\ref{eqn:potential_fromzero}),
$\left[\mu_i(T,p) - \mu_i^0\right]$, requires a reference state of zero, while
the majority of data in the literature is relative to standard conditions of
298.15 K and 1 bar. It is therefore convenient to break up the correction to use
this reference state, which is denoted with a superscript $\theta$:
\begin{align}
\mu_i(T,p_i) &= E_i^\textrm{DFT} + E_i^\textrm{ZP} + 
		\left[\mu_i(T,p_i) - \mu_i^\theta\right] + 
		\left[\mu_i^\theta - \mu_i^0 \right] \label{eqn:potential_split}\\
\intertext{Introducing the relationship with enthalpy, $\mu_i = H_i - TS$:}
\mu_i^\theta - \mu_i^0 &= \left[ H_i^\theta - (TS)^\theta \right] - 
                          \left[ H_i^0 - (TS)^0 \right] \\
\intertext{$(TS)^0 = 0$, so simplifying and rearranging:}
\mu_i^\theta - \mu_i^0  &= \left[H_i^\theta - H_i^0 \right] - (TS)^\theta
\end{align}
\begin{widetext}
\begin{align}
\intertext{Substituting this back into Eq. (\ref{eqn:potential_split}):}
\mu_i(T,p_i) &= E_i^\textrm{DFT} + E_i^\textrm{ZP} + \left[\mu_i(T,p_i) -
 \mu_i^\theta\right] + \left[H_i^\theta - H_i^0 \right] - (TS)^\theta
\label{eqn:potential_long}
\\
\intertext{Of these terms: 
$E_{i}^{\text{DFT}}$ is found by \emph{ab initio} calculations;
$E_{i}^{\text{ZP}}$ depends on the lattice frequency and is relatively small (literature values are given in Table~\ref{table:gas-zpe});
$\left[H_{i}^{\theta} - H_{i}^{0} \right]$ and 
$(TS)^{\theta} = 298.15 \text{K}\times S^{\theta}$
are available from literature data.
The only variable term is $\left[\mu_i(T,p_i) - \mu_i^\theta\right]$. Introducing the group $\mu_i(T,p_i^{\theta})$ to break the
process into isothermal pressure change and isobaric temperature change:}
\left[\mu_i(T,p_i) - \mu_i^\theta\right] &= 
	\left[ \mu_i(T,p_i) - \mu_{i}(T,p_{i}^{\theta})\right] + 
	\left[ \mu_{i}(T,p_{i}^{\theta}) - \mu_{i}^{\theta} \right]
\\
\intertext{To account for the pressure change, we use an ideal-gas relationship:\cite{Warn1996}}
\left[\mu_{i}(T,p_{i}) - \mu_i(T,p_{i}^{\theta})\right] &= 
	RT \ln (p_{i}/p_{i}^\theta)  \\
\left[\mu_i(T,p_{i}) - \mu_i^\theta \right] &= 
	RT \ln (p_{i}/p_{i}^\theta) + \left[ \mu_i(T,p_{i}^\theta) -
	 \mu_i^\theta\right] \label{eqn:pressure_correction}
\\
\intertext{The temperature change uses the standard constant-pressure heat capacity
$C_{p}(T) = \left(\frac{\partial H}{\partial T}\right)_{p^{\theta}}$:}
\left[ \mu_{i}(T,p_{i}^{\theta}) - \mu_{i}^{\theta} \right] &= 
	\left[H_i(T,p_{i}^\theta) - H_i^\theta \right] - 
	\left[ TS(T,p_{i}^{\theta}) - T^{\theta} S^{\theta} \right] 
\\
\left[ \mu_i(T,p_{i}^\theta) - \mu_i^\theta\right] &=
	\int^T_{T^{\theta}}C_{p}\text{d}T - \left[ TS(T,p_{i}^{\theta}) -
	 T^{\theta} S^\theta \right] \label{eqn:temperature_correction}
\\
\intertext{Combining Equations \eqref{eqn:potential_long},
 \eqref{eqn:pressure_correction} and \eqref{eqn:temperature_correction}:}
\mu_i(T,p_i) &= E^\textrm{DFT} + E^\textrm{ZP} + RT \ln (p_i/p_i^\theta) + \int^T_{T^\theta}C_p\textrm{d}T - \left[ TS(T,p_i^\theta) - T^\theta S^\theta \right]
+ \left[H_i^\theta - H_i^0 \right] - (TS)^\theta \\
\mu_i(T,p_i) &= \underbrace{E^\textrm{DFT} + E^\textrm{ZP}}_\text{Energy at zero} + \underbrace{\left[H_i^\theta - H_i^0 \right]}_\text{Standard enthalpy} + \underbrace{\int^T_{T^\theta}C_p\textrm{d}T}_\text{Enthalpy correction}  + \underbrace{RT \ln (p_i/p_i^\theta)}_\text{Free energy correction} -  \underbrace{TS(T,p_i^\theta)}_\text{Entropic contribution}  \label{eqn:gas_mu}
\end{align}
\end{widetext}
The key pieces of data needed are therefore the standard enthalpy
$\left[H_i^\theta - H_i^0 \right]$, the heat capacity, $C_p$, and entropy, $S$,
as functions of temperature at standard pressure. The standard enthalpy is
available from reference books, while the temperature-dependant heat capacity and entropy are obtained
from tables or polynomial equations as discussed above.

\subsection{Lattice dynamics}
\label{sec:phonon-method}
Thermal properties were calculated within the harmonic approximation
using the {\sc Phonopy 1.5} software package, preparing and post-processing
{\sc FHI-aims} calculations.\cite{*[{}] [{ [{\sc Phonopy} is available as an open-source package from http://phonopy.sourceforge.net] }] Togo2008}
The number of \kpoints{} was scaled to match the 10 \AA{} target length
cutoff employed in relaxation calculations. Forces were calculated for
atomic displacements of 0.01 \AA{}, with a convergence threshold of
$1\times10^{-5}$ eV \AA{}$^{-1}$.

{\sc Phonopy} follows the Parlinski-Li-Kawazoe method to generate a ``dynamical matrix'' of forces describing the harmonic behaviour of the atoms in the system. In this scheme the second derivatives of energy are obtained by combining analytical first derivatives with small displacements in supercells.\cite{Parlinski1997}
These second derivatives yield a set of phonon frequencies, $\omega$, 
which may be expressed as a phonon band structure and density of states (DOS).\cite{Stoffel2010}
Ultimately a thermodynamic partition function can be formed for each mode, 
\begin{equation}
Z_i = \sum\limits_j\exp\left(\frac{E_j}{k_B T}\right),
\end{equation}
where the energy at a given state $E_j = \hbar \omega_j$; the product of all $Z_i$ yields an overall partition function $Z$ from which the Helmholtz free energy
\begin{equation}
A = -k_B T ln Z .
\end{equation}
This energy includes the zero-point energy at 0 K, and by differentiation the heat capacity, entropy and related properties are obtained as functions of temperature.
In this study the relationship was sampled over temperature with a density of at least one
point per degree Kelvin, and any intermediate values were estimated
using cubic spline interpolation.

It is possible to use the quasi-harmonic approximation to account for the
influence of pressure and thermal expansion by modelling compressed
and expanded supercells; \ce{GaN} is relatively incompressible, with a bulk modulus of over 200 GPa, and in this study the effect is assumed to be
negligible.\cite{Sarasamak2010} 
A higher level of accuracy can be obtained by the use of molecular dynamics (MD) simulations with an appropriate thermostat.
The computational cost of this approach is very high, as many time steps are needed to obtain a converged statistical average.
Nonetheless the approach accounts for higher-order anharmonicity, within the scope of the analytical potentials or \emph{ab initio} method used to calculate energies and forces.

Given a suitable function for heat capacity, it is possible to expand Eq.~(\ref{eqn:potential_fromzero}) and calculate the chemical potentials of known crystals under an absolute pressure $P$:
\begin{equation}
\mu_{i}(T,P) = E_i^{\text{DFT}} + E_i^{\text{ZP}} + \int^T_0 C_p \text{d}T + PV - TS_\text{vib}(T)
\end{equation}
where $E^\text{ZP}$, $C_p(T)$ and the vibrational entropy contribution $S_\text{vib}(T)$ are drawn from the dynamic lattice model, and $PV$ is simply the total pressure multiplied by the specific volume. 

\subsection{Derivation of free energy}

The materials are assumed to be ideal in that the chemical potentials, $\mu_i$
of each component $i$ in isolation may simply be summed together to obtain the
overall Gibbs free energy of reaction, $\Delta G_r$:
\begin{equation}\label{eqn:sum_potentials}
\Delta G_{r}(T,P) = \sum\limits_i \mu_{i}(T,p)\Delta n_i
\end{equation}
where $T$ is the temperature, $P$ is the total system pressure, $p$ is the
partial pressure of a component and $\Delta n_i$ is the
stoichiometric change for the component $i$. 
A more advanced model might substitute the partial pressures for fugacities.

\section{results}
\subsection{Bulk thermodynamic properties}
\subsubsection{Solid-state thermodynamic potentials}
\begin{figure}
\includegraphics[scale=1]{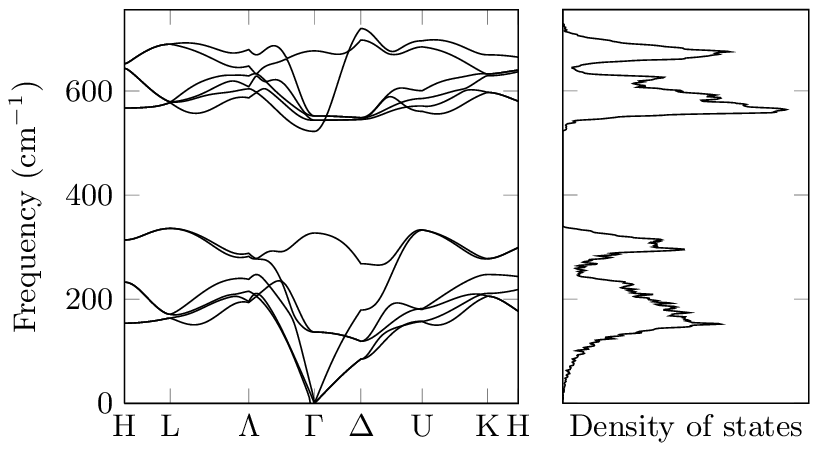}
\caption{Phonon band structure and density of states for GaN
\label{fig:GaN_phonondos}}
\end{figure}

\begin{figure}
\includegraphics[scale=1]{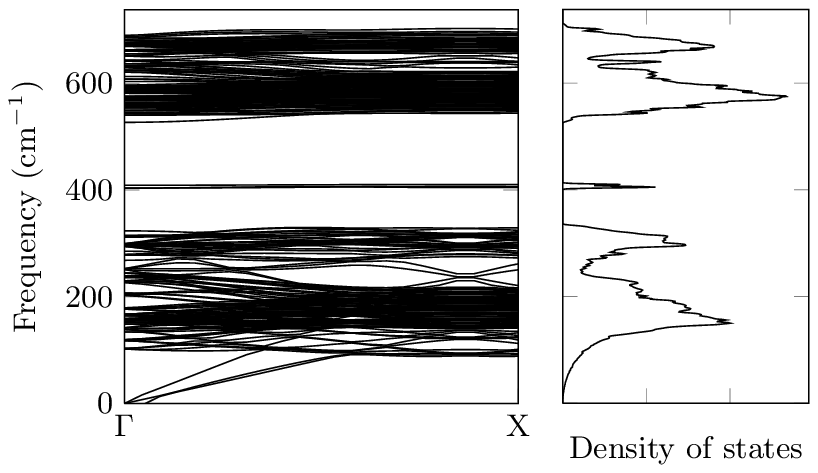}
\caption{Phonon band structure and density of states for \ce{Ga36N35O}
\label{fig:S_72_phonondos}}
\end{figure}

\begin{figure}
\includegraphics[scale=1]{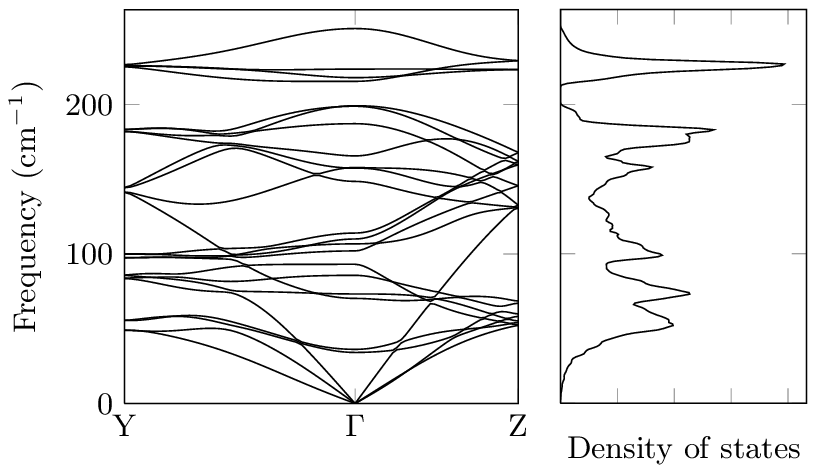}
\caption{Phonon band structure and density of states for Ga
\label{fig:Ga_phonondos}}
\end{figure}

\begin{figure}
\includegraphics[scale=1]{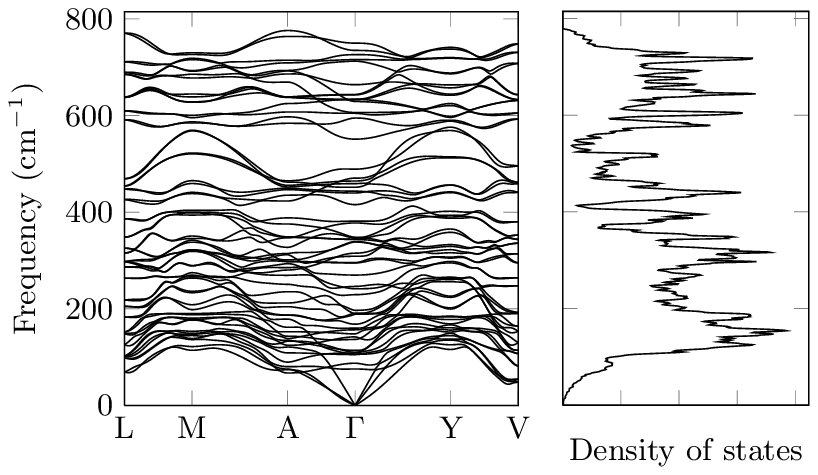}
\caption{Phonon band structure and density of states for \ce{Ga2O3}
\label{fig:Ga2O3_phonondos}}
\end{figure}

\begin{figure}
\includegraphics[scale=1]{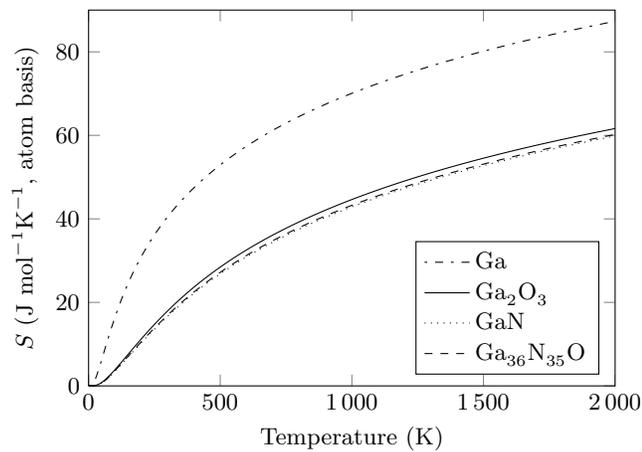}
\caption{Solid-state vibrational entropies from phonon calculations
\label{fig:solid_S}}
\end{figure}

The lattice dynamic calculations described in section
\ref{sec:phonon-method} were applied to relaxed structures of Ga, GaN,
\ce{Ga2O3} to obtain free energies, entropies and heat capacities.
 The dispersion curve and DOS for GaN shows similar behaviour to work based on analytical potential models and Raman
 spectroscopy\cite{Azuhata1996,Siegle1997,Davydov1998}, while the \ce{Ga2O3} curves may be compared to a previous \emph{ab initio} study.\cite{Liu2007}
Two defect-containing supercells were also subjected to this analysis: the
72- and 128-atom supercells in which one nitrogen atom is substituted
for oxygen, \ce{Ga36N35O} and \ce{Ga64N63O}. The calculated zero-point
energies are given with standard-temperature Helmholtz free energies ($A = U-TS$)
and heat capacities in Table~\ref{table:thermal_properties}, 
and phonon band structures in Figures~\ref{fig:GaN_phonondos}-\ref{fig:Ga2O3_phonondos}.
\begin{table*}
\setlength{\extrarowheight}{2pt}
\caption{Thermal properties from phonon calculations with {\sc FHI-aims} and {\sc Phonopy}: Zero-point energy $E^\text{ZP}$; standard Helmholtz free energy $A^{298.15\text{K}}$; standard heat capacity $C_p^{298.15\text{K}}$.
\label{table:thermal_properties}
}
\begin{ruledtabular}
\begin{tabular}{l c c d d d c d d d}
\multirow{3}{*}{Compound} & \multirow{3}{*}{Supercell} & 
	\multirow{3}{*}{\kpoints{}}  & 
    \multicolumn{3}{c}{Unit cell basis} && \multicolumn{3}{c}{Formula unit
      basis}\\
& & &
    \multicolumn{1}{c}{$E^{\text{ZP}}$} & 
    \multicolumn{1}{c}{$A^{298.15 \text{K}}$} & 
    \multicolumn{1}{c}{$C_{p}^{298.15 \text{K}}$} &&
    \multicolumn{1}{c}{$E^{\text{ZP}}$} & 
    \multicolumn{1}{c}{$A^{298.15 \text{K}}$} & 
    \multicolumn{1}{c}{$C_{p}^{298.15 \text{K}}$} 
\\ 
& & & 
    \multicolumn{1}{c}{$/\text{eV}$} & 
    \multicolumn{1}{c}{$/\text{eV}$} & 
    \multicolumn{1}{c}{$/k_B$} &&
    \multicolumn{1}{c}{$/\text{kJ mol}^{-1}$} & 
    \multicolumn{1}{c}{$/\text{kJ mol}^{-1}$} & 
    \multicolumn{1}{c}{$/\text{J mol$^{-1}$K$^{-1}$}$}
\\ \hline 
GaN & [  3 3 2 ] & [3 3 3 ] & 0.305 & 0.217 & 8.529&& 14.70 & 10.45 & 35.46 \\ 
Ga & [  2 2 2 ] & [3 2 3 ] & 0.194 & -0.356 & 23.069&& 2.34 & -4.29 & 23.98 \\ 
\ce{Ga2O3} & [  1 3 2 ] & [2 3 2 ] & 1.412 & 0.933 & 44.597&& 34.07 & 22.50 & 92.70 \\ 
\ce{Ga36N35O} & [  3 3 2 ] & [1 1 1 ] & 5.410 & 3.791 & 154.715&& 521.94 & 365.76 & 1286.37
\end{tabular}
\end{ruledtabular}
\end{table*}
The temperature variation of entropy is presented in Figure~\ref{fig:solid_S}: on an atomic basis, the values for
the oxygen and nitrogen compounds are relatively close, while the Ga
metal has greater entropy. The overall defect entropy change appears
to be of the order 25 J mol$^{-1}$ K$^{-1}$; this is comparable with other work on point defects in ionic compounds, and corresponds to a non-negligible amount of energy at high temperatures.\cite{Gillan1983,Walsh2011a}

\subsubsection{Heat capacities}
\begin{figure}
\includegraphics[scale=1]{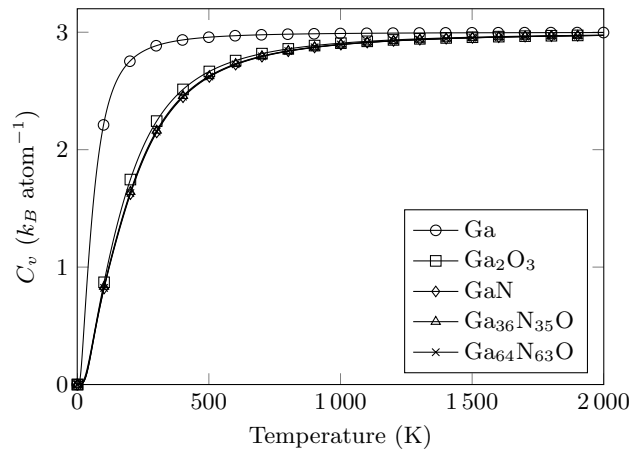}
\caption{Solid-state heat capacities from phonon calculations
\label{fig:solid_cp}}
\end{figure}

\begin{figure}
\includegraphics{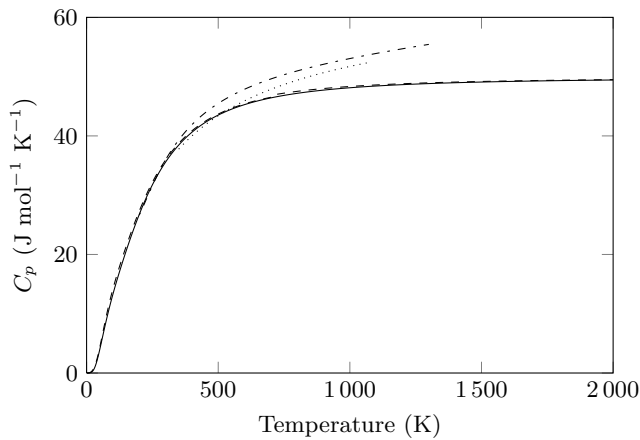}
\caption{Heat capacity for GaN: (--) PBEsol/harmonic approximation;
  (-~-) Debye/Einstein fit to adiabatic calorimetry measurements,
  Danilchenko \emph{et al.} (2006)\cite{Danilchenko2006}; ($\cdots$) DSC
  measurements and empirical fit by Jacob \emph{et al.} (2007)
  \cite{Jacob2007}; (-~$\cdot$~-) Fit by Leitner \emph{et al.} (2003),
  incorporating data from Calvet calorimetry and other
  literature.\cite{Leitner2003}
\label{fig:gan_cp}}
\end{figure}

All computed heat capacities are given over the studied temperature range in
Figure~\ref{fig:solid_cp}; the behaviour of the gallium compounds is
extremely close, and all materials tend towards the Dulong-Petit limit
of $3k_B$ per atom.
Comparing these results with the literature, the \emph{ab initio} heat capacity of GaN is plotted against several
fits to experimental data in Figure~\ref{fig:gan_cp}. Danilchenko et
al. fitted a mixed Debye and Einstein model to data from
low-temperature calorimetry, while Leitner \emph{et al.} carried out
high-temperature measurements and formed an empirical model including
data from other researchers.\cite{Leitner2003} Jacob \emph{et al.} (2007)
used differential scanning calorimetry (DSC) to return a slightly lower set of high-temperature heat
capacity data.\cite{Jacob2007} Not shown is the result of Sanati and
Estreicher's \emph{ab initio} calculation, which used a  Ceperley-Alder local-density
functional at a single \kpoint{}, and appears to
give similar results to ours.\cite{Sanati2004} Of interest is the fact
that the systems based on theory (phonon integration and
Debye/Einstein models) tend towards the Dulong-Petit limit, while the experimental/empirical work exceeds this limit. The simple
harmonic approaches do not account for electronic and anharmonic
contributions, which may yield this additional heat capacity.
In addition to the deviations in heat capacity, it is worth bearing in mind that Ga is molten above around 300 K and simple thermal decomposition of GaN is not taken into account: these are reference states rather than physical models.

\begin{figure}
\includegraphics{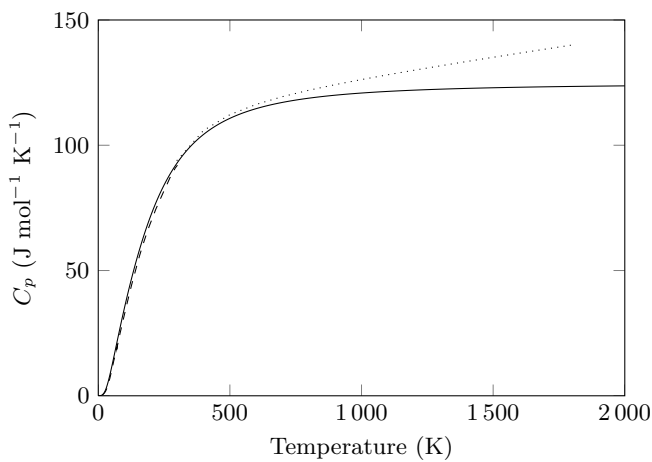}
\caption{Heat capacity for \ce{Ga2O3}: (--) PBEsol/harmonic approximation;
  (-~-) Low temperature calorimetry data from Adams (1952)\cite{AdamsJr.1952}; ($\cdots$) High temperature calorimetry from Mills (2007).\cite{Mills1972}
\label{fig:cp_galox}}
\end{figure}

\begin{figure}
\includegraphics[scale=1]{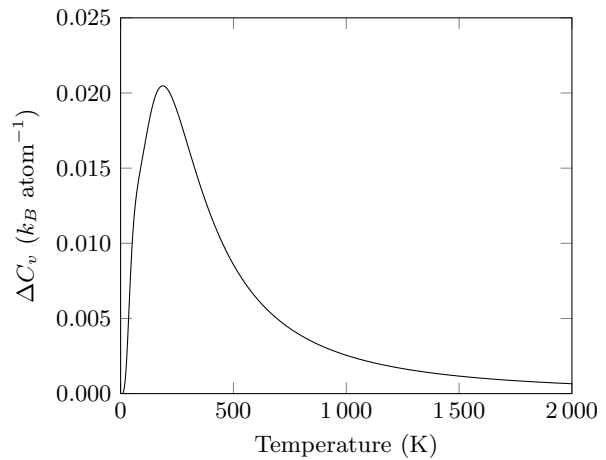}
\caption{Difference in heat capacity between pure \ce{GaN} and 72-atom supercell with single oxygen substitution:
$\Delta C_v = C_{v, \textrm{Ga$_{36}$N$_{35}$O}} - C_{v, 36\textrm{GaN}}$
\label{fig:cp_diff}}
\end{figure}

The heat capacity of \ce{Ga2O3} shows similar behaviour relative to the literature, with a close correspondence at low temperatures and deviation from around 500 K (Figure~\ref{fig:cp_galox}).
The phonon band structures for GaN
(Figures~\ref{fig:GaN_phonondos} and \ref{fig:S_72_phonondos}) clearly show the
impact of the defect; a cluster of three `gap bands' appears at around
400~cm$^{-1}$, providing a set of vibrational energies which are
available at lower temperatures than the optic modes above
500~cm$^{-1}$.
These correspond to highly-localised vibrations of the substitutional oxygen atom.
The effect manifests itself as a peak in the difference in
heat capacity of pure and oxygen-doped \ce{GaN}
(Figure~\ref{fig:cp_diff}). The impact of the defect is both
qualitatively logical and quantitatively negligible. 
Given the comparatively small contribution of the heat capacity to the free energy,
it should generally be an acceptable approximation to use the 
heat capacity of the host material in thermodynamic models.

\subsubsection{Enthalpy of formation}
Enthalpies of formation at standard conditions were calculated using a simplified form of Equations~(\ref{eqn:gas_mu}) and~(\ref{eqn:sum_potentials}) given that $H=G+TS$:

\begin{equation}
\Delta H_{f}^{\theta} = \sum\limits_i \left(E^{\text{DFT}} + E^{\text{ZPE}} +
						\left[H^{\theta} - H^{0}\right] \right) \Delta n_{i}
\end{equation}

\begin{figure}
\includegraphics[scale=1]{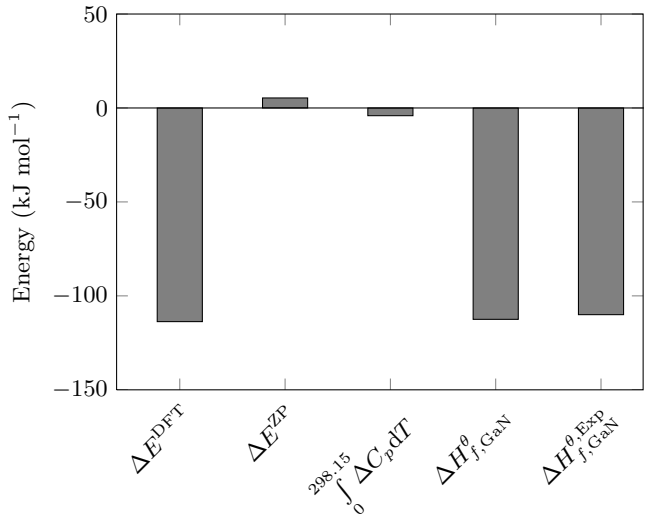}
\caption{Contribution of energy terms to overall \emph{ab initio} formation enthalpy of \ce{GaN} ($\Delta H^{\theta}_{f,\ce{GaN}}$) compared to experimental values $\Delta H^{\theta,\text{CRC}}_{f,\ce{GaN}}$ and $\Delta H^{\theta,\text{2009}}_{f,\ce{GaN}}$, from the CRC handbook and Jacob and Rajita (2009), respectively.\cite{handbook-crc, Jacob2009}
\label{fig:gan-enthalpy}}
\end{figure}

\begin{figure}
\includegraphics[scale=1]{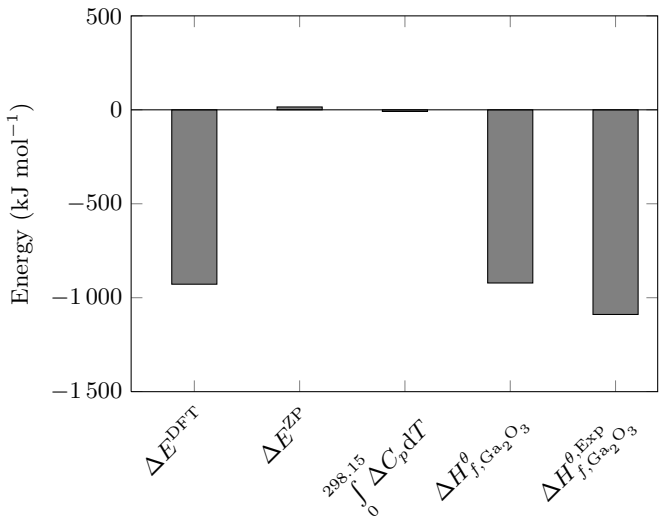}
\caption{Contribution of energy terms to overall \emph{ab initio} formation enthalpy of \ce{Ga2O3} ($\Delta H^{\theta}_{f,\ce{Ga2O3}}$) compared to experimental value $\Delta H^{\theta,\text{Exp}}_{f,\ce{Ga2O3}}$.\cite{handbook-crc}
\label{fig:galox-enthalpy}}
\end{figure}

The resulting values are given in Table~\ref{table:enthalpies} and
compared to classic experimental values. While the value for \ce{GaN}
agrees with the literature to within a few kJ mol$^{-1}$, there is a
greater discrepancy for \ce{Ga2O3}.  As seen in Figures
\ref{fig:gan-enthalpy} and \ref{fig:galox-enthalpy}, the overall formation enthalpy is dominated by
the ground-state potential energy; the common approximation of
comparing ground state energies to standard enthalpy changes could be
justified in this case.
\begin{table}
\caption{Predicted and experimental enthalpies of formation\cite{handbook-crc}
\label{table:enthalpies}}
\begin{ruledtabular}
\begin{tabular}{c d d}
\multirow{2}{*}{Material} & \multicolumn{1}{c}{$\Delta H^\text{calculated}$} & 
		   \multicolumn{1}{c}{$\Delta H^\text{experimental}$} \\
 &  \multicolumn{1}{c}{/ kJ mol$^{-1}$} &
 \multicolumn{1}{c}{/ kJ mol$^{-1}$} \\
\hline
\ce{GaN}    & -112.49       & -110    \\
\ce{Ga2O3}  & -921.64       & -1089.1     \\
\end{tabular}
\end{ruledtabular}
\end{table}

While the agreement between this work and the established literature
is close, it is worth observing that the standard enthalpy of
formation of GaN has been the subject of some debate in recent years:
Jacob \emph{et al.} suggested a value of -126.792 kJ mol$^{-1}$ in 2007,
Peshek \emph{et al.} obtained -165 kJ mol$^{-1}$ in 2008 and Jacob and
Rajitha responded in 2009 with a critical letter and a new value of
-129.289 kJ mol$^{-1}$.\cite{Jacob2007,Peshek2008,Jacob2009} All of these are greater in magnitude than the value of -110 kJ mol $^{-1}$ reported in the current CRC Handbook of Chemistry and Physics.\cite{handbook-crc}

\subsection{Complete oxidation}
\begin{figure}
\includegraphics[scale=1]{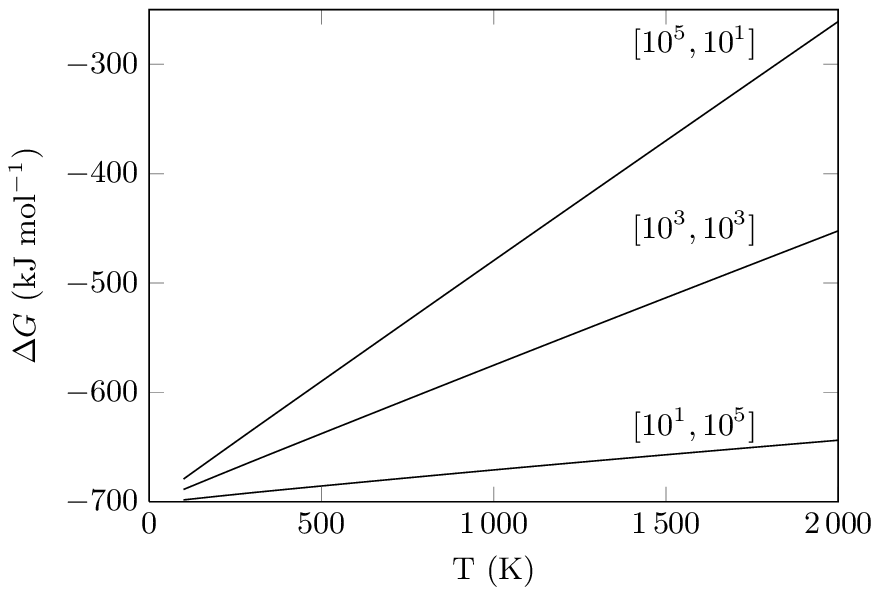}
\caption{Modelled Gibbs free energy of complete oxidation from GaN to \ce{Ga2O3}. For each line the corresponding partial pressures, $[p_\ce{N2},p_\ce{O2}]$, are given in Pa.
\label{fig:full_ox}
}
\end{figure}

\begin{figure}
\includegraphics[scale=1]{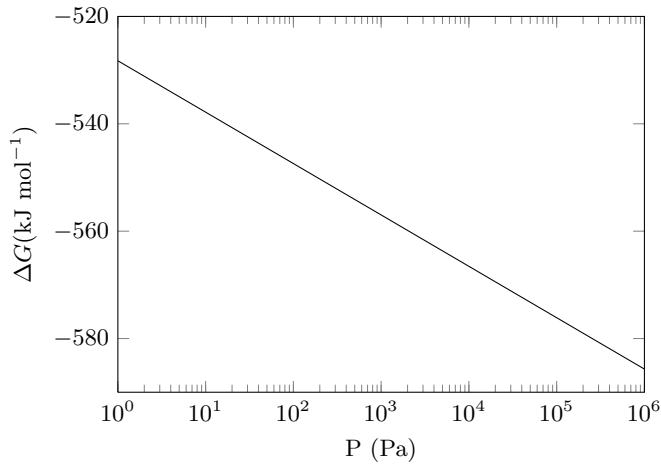}
\caption{Modelled Gibbs free energy of complete oxidation from GaN to \ce{Ga2O3} at 1000 K. Absolute pressure is varied for an atmosphere containing 20\%$_\text{vol}$~\ce{O2}, 80\%$_\text{vol}$~\ce{N2}.
\label{fig:full_oxidation_P}
}
\end{figure}

The calculated Gibbs free energy of the oxidation from GaN to \ce{Ga2O3} is given
for a wide range of conditions in Figure~\ref{fig:full_ox}. At standard temperature and pressure, with an air-like \ce{O2}:\ce{N2} pressure ratio of 20:80, the predicted Gibbs free energy of oxidation is -663.5 kJ mol$^{-1}$.
There is a mild effect of absolute pressure, dropping the energy further by tens of kJ mol$^{-1}$ as demonstrated in Figure~\ref{fig:full_oxidation_P}, while the relative ratio of gases has a dramatic effect of the order 100~kJ~mol$^{-1}$ at elevated temperature (Figure~\ref{fig:full_ox}).
This is because the entropy of the gas phase decreases at high pressures, and it becomes less unfavourable to remove material to form a solid.
At very high pressures (of the order GPa, not shown here) the trend is reversed, but this is beyond the scope of the model as it is driven by the relative compressibilities of gases and solids. 
The strongly negative $\Delta G$ values indicate that oxidation is favourable at equilibrium under
all industrially-feasible reaction conditions; the GaN-air system,
while temperature-resistant in practice, is not thermodynamically stable. This implies that it relies entirely on kinetic stability.

\subsection{Dilute oxidation}
\begin{figure}
\includegraphics[scale=1]{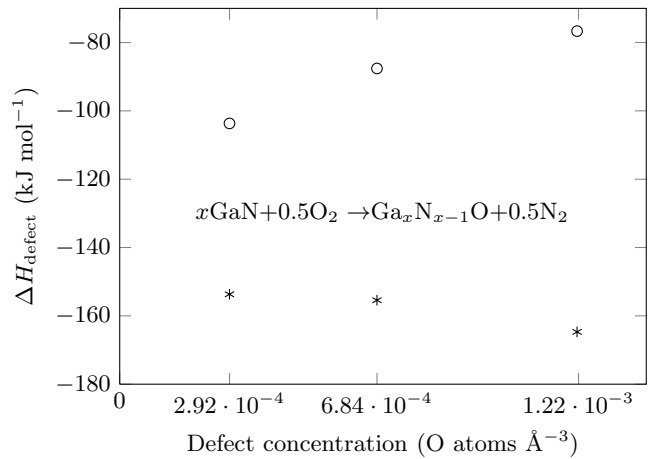}
\caption{
$\circ$
: Enthalpy change (298.15 K, 0.2 bar \ce{O2}, 0.8~bar \ce{N2}) associated with partial oxidation (single oxygen substitution for GaN supercells);
$*$:
enthalpy change including band filling correction (extrapolation to dilute limit).
\label{fig:conc_dependence_H}}
\end{figure}

The enthalpies of dilute oxidation (i.e. substitutional oxygen defect
formation) can be explored using the energies of large supercells,
taking advantage of the relatively small deviation in heat capacity
from the pure substance. In Figure~\ref{fig:conc_dependence_H}, defect
formation enthalpies are calculated for standard conditions in an
air-like mixture as:
\begin{equation}
\Delta H_\text{defect} = H_{\text{Ga}_{x}\text{N}_{x-1}\text{O}} + 0.5 H_\ce{N2} - x H_\ce{GaN} - 0.5 H_\ce{O2}
\end{equation}
and the vibrational contributions to the two solids are assumed to be equivalent as the difference in heat capacity is of the order 0.05~J~mol$^{-1}$~K$^{-1}$ (Figure~\ref{fig:cp_diff}). 

The overall defect formation enthalpy is exothermic for all concentrations studied; however, there is a strong dependence on the supercell size, which is varied from \ce{Ga36N36} to \ce{Ga150N150}.
Application of the band-filling correction further stabilises partial oxidation by $\sim{}40$~kJ~mol$^{-1}$, and the values for the two larger supercells appear to approach convergence. 

\begin{figure}
\includegraphics[scale=1]{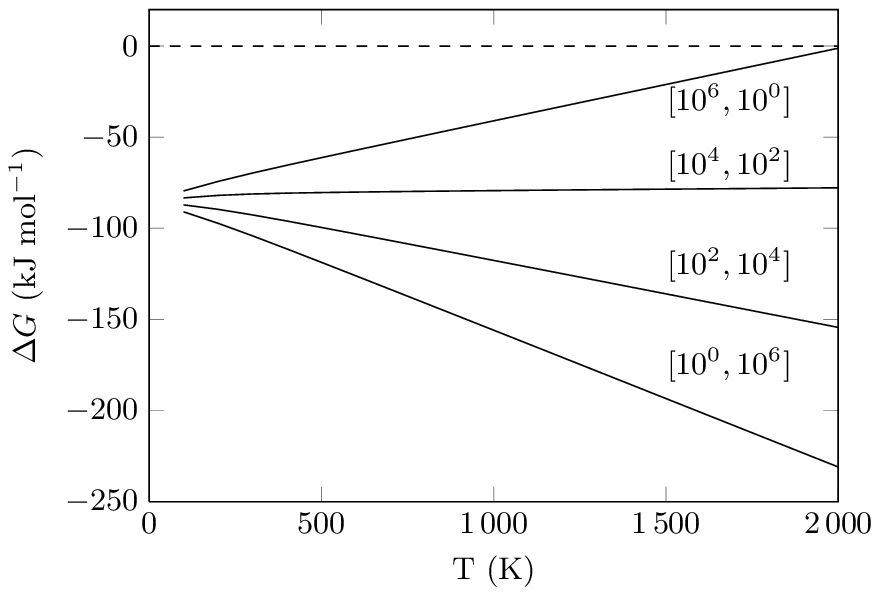}
\caption{Modelled Gibbs free energy of partial oxidation from GaN to \ce{Ga36N35O}. For each line the corresponding partial pressures, $[p_\ce{N2},p_\ce{O2}]$, are given in Pa.
\label{fig:S_72_sub}
}
\end{figure}

\begin{figure}
\includegraphics[scale=1]{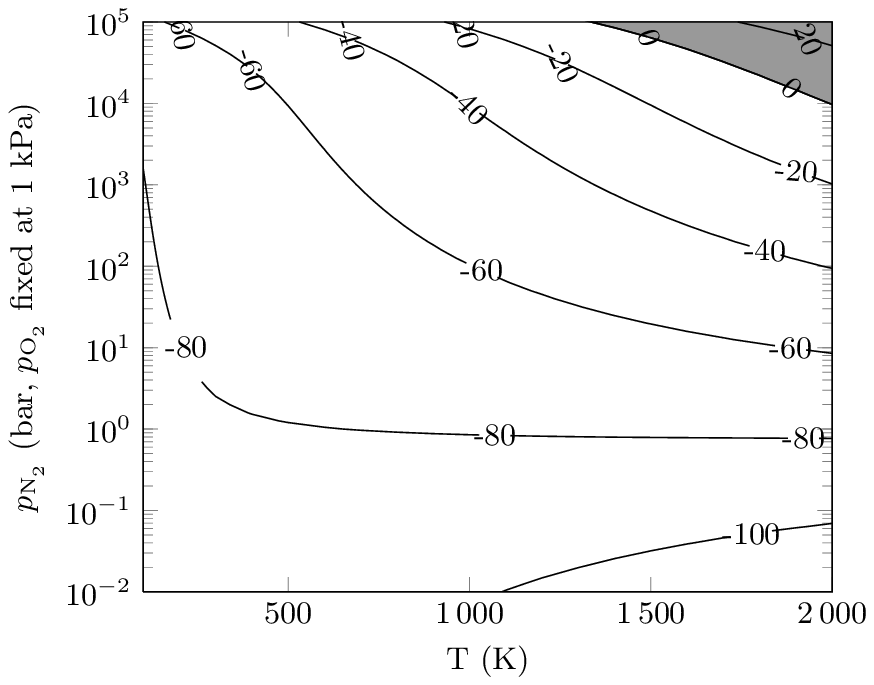}
\caption{Gibbs free energy surface of oxygen defect formation in a 72-atom GaN supercell. Contours are labelled with Gibbs free energy change ($\Delta G$ in kJ mol$^{-1}$); values above zero (i.e. unfavourable) are shaded.
\label{fig:G_contours_72}}
\end{figure}

\begin{figure}
\includegraphics[scale=1]{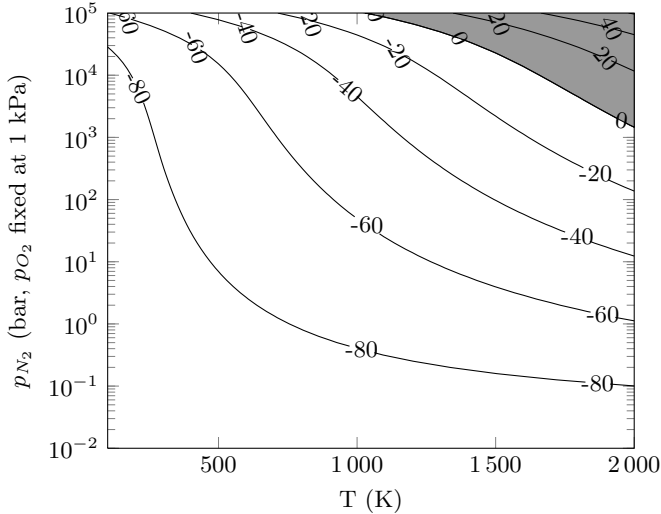}
\caption{Gibbs free energy surface of oxygen defect formation in a 128-atom GaN supercell. Contours are labelled with Gibbs free energy change ($\Delta G$ in kJ mol$^{-1}$); values above zero (i.e. unfavourable) are shaded.
\label{fig:G_contours_128}}
\end{figure}

The modelled Gibbs free energy for dilute oxidation is given in Figure~\ref{fig:S_72_sub}
for a single substitution per 72-atom cell over a range of
temperatures and partial pressures. At the most extreme conditions the
threshold of $\Delta G=0$ is approached; the operating envelope may be
considered more easily as a contour map as in
Figures~\ref{fig:G_contours_72} and \ref{fig:G_contours_128}. Comparing
these envelopes for two defect concentrations, we observe that it is
actually easier to achieve a positive $\Delta G$ for the more dilute
defect. The contribution of entropy to the free
energy reverses the trend seen for enthalpy in Figure~\ref{fig:conc_dependence_H}. 
For example, in the high-pressure high-temperature regime (2000 K, $10^4$ bar), the ground state ($\Delta E^\text{DFT}$) contributions to Gibbs free energy are -80.7 kJ mol$^{-1}$ and -91.6 kJ mol$^{-1}$ for the 72-atom and 128-atom supercells, respectively.
In the same region, the entropic $(-RT\ln(p_i/p_i^\theta{})-TS)$ contributions are -305.6 kJ mol$^{-1}$ and -278.0 kJ mol$^{-1}$.
The enthalpy corrections $(\int{}C_p\text{d}T + PV)$ of 117.8 kJ mol$^{-1}$ and 119.2 kJ mol$^{-1}$ show a smaller concentration dependence.

\section{Conclusions}
A thermodynamic model has been developed for the GaN-\ce{O2}-\ce{N2}
system from \emph{ab initio} calculations and readily-available
thermodynamic data. The model permits the free energies and enthalpies
of oxidation and defect formation to be calculated for any conditions
within a practical processing range. The case of complete oxidation to \ce{Ga2O3} represents an overall driving force, and the behaviour following a phase transition, while defect calculations offer insight to the early onset of oxidation.

Varying the ratio of gases provides a strong entropic driving force in
the system, and industrial processing conditions are capable of
shifting the equilibrium. However, the oxidation of GaN appears to be
such a favourable reaction that in practice extreme conditions would be
required to prevent oxides or substitutional defects from
being thermodynamically stable. 
The high thermal stability of GaN with respect to oxygen is
therefore kinetic in nature.

\appendix
\section{Supercell transformations
\label{sec:transform}}

The transformation matrices in Table~\ref{table:transf} were applied
with the VESTA software package\cite{Momma2011} to form supercells from the relaxed GaN unit cell.

\begin{table}
\setlength{\extrarowheight}{5pt}
\caption{Generation of supercells from GaN unit cell}
\label{table:transf}
\begin{ruledtabular}
\begin{tabular}{c c c}
Atoms in cell & Shape & Transformation matrix \\ \hline
4 & Hexagonal &    $\left( \begin{array}{c c c}
		1 & 0 & 0 \\
		0 & 1 & 0 \\
		0 & 0 & 1
		\end{array} \right)$ \\
72 & Hexagonal &   $\left( \begin{array}{c c c}
		3 & 0 & 0 \\
		0 & 3 & 0 \\
		0 & 0 & 2
\end{array}\right)$ \\
128 & Orthorhombic &$\left( \begin{array}{c c c}
		4 & 0 & 0 \\
		2 & 4 & 0 \\
		0 & 0 & 2
\end{array}\right)$ \\
300 & Hexagonal &  $\left( \begin{array}{c c c}
		5 & 0 & 0 \\
		0 & 5 & 0 \\
		0 & 0 & 3
\end{array}\right)$
\end{tabular}
\end{ruledtabular}
\end{table}

\begin{acknowledgments}
We thank D. Allsopp for useful discussions.
We acknowledge the use of the Chemical Database Service at Daresbury and the Inorganic Crystal Structure Database (ICSD).
This work was funded and supported by the EPSRC through the Doctoral Training Centre in Sustainable Chemical Technologies at the University of Bath (EP/G03768X/1).
Via our membership of the UK's HPC Materials Chemistry Consortium, which is funded by EPSRC (EP/F067496), this work made use of the facilities of HECToR, the UK's national high-performance computing service, which is provided by UoE HPCx Ltd at the University of Edinburgh, Cray Inc and NAG Ltd, and funded by the Office of Science and Technology through EPSRC's High End Computing Programme.
Large structure relaxations and lattice dynamics were calculated using Blue Joule, a Bluegene/Q system at the Science and Technology Facility Council's Daresbury Laboratory, which we were kindly permitted to use as part of the Early Access program.
\end{acknowledgments}


\begin{thebibliography}{52}%
\makeatletter
\providecommand \@ifxundefined [1]{%
 \@ifx{#1\undefined}
}%
\providecommand \@ifnum [1]{%
 \ifnum #1\expandafter \@firstoftwo
 \else \expandafter \@secondoftwo
 \fi
}%
\providecommand \@ifx [1]{%
 \ifx #1\expandafter \@firstoftwo
 \else \expandafter \@secondoftwo
 \fi
}%
\providecommand \natexlab [1]{#1}%
\providecommand \enquote  [1]{``#1''}%
\providecommand \bibnamefont  [1]{#1}%
\providecommand \bibfnamefont [1]{#1}%
\providecommand \citenamefont [1]{#1}%
\providecommand \href@noop [0]{\@secondoftwo}%
\providecommand \href [0]{\begingroup \@sanitize@url \@href}%
\providecommand \@href[1]{\@@startlink{#1}\@@href}%
\providecommand \@@href[1]{\endgroup#1\@@endlink}%
\providecommand \@sanitize@url [0]{\catcode `\\12\catcode `\$12\catcode
  `\&12\catcode `\#12\catcode `\^12\catcode `\_12\catcode `\%12\relax}%
\providecommand \@@startlink[1]{}%
\providecommand \@@endlink[0]{}%
\providecommand \url  [0]{\begingroup\@sanitize@url \@url }%
\providecommand \@url [1]{\endgroup\@href {#1}{\urlprefix }}%
\providecommand \urlprefix  [0]{URL }%
\providecommand \Eprint [0]{\href }%
\providecommand \doibase [0]{http://dx.doi.org/}%
\providecommand \selectlanguage [0]{\@gobble}%
\providecommand \bibinfo  [0]{\@secondoftwo}%
\providecommand \bibfield  [0]{\@secondoftwo}%
\providecommand \translation [1]{[#1]}%
\providecommand \BibitemOpen [0]{}%
\providecommand \bibitemStop [0]{}%
\providecommand \bibitemNoStop [0]{.\EOS\space}%
\providecommand \EOS [0]{\spacefactor3000\relax}%
\providecommand \BibitemShut  [1]{\csname bibitem#1\endcsname}%
\let\auto@bib@innerbib\@empty
\bibitem [{\citenamefont {Murphy}(2012)}]{Murphy2012}%
  \BibitemOpen
  \bibfield  {author} {\bibinfo {author} {\bibfnamefont {T.~W.}\ \bibnamefont
  {Murphy}},\ }\href {\doibase 10.1063/1.4721897} {\bibfield  {journal}
  {\bibinfo  {journal} {J. Appl. Phys.}\ }\textbf {\bibinfo {volume} {111}},\
  \bibinfo {pages} {104909} (\bibinfo {year} {2012})}\BibitemShut {NoStop}%
\bibitem [{\citenamefont {Narukawa}\ \emph {et~al.}(2010)\citenamefont
  {Narukawa}, \citenamefont {Ichikawa}, \citenamefont {Sanga}, \citenamefont
  {Sano},\ and\ \citenamefont {Mukai}}]{Narukawa2010}%
  \BibitemOpen
  \bibfield  {author} {\bibinfo {author} {\bibfnamefont {Y.}~\bibnamefont
  {Narukawa}}, \bibinfo {author} {\bibfnamefont {M.}~\bibnamefont {Ichikawa}},
  \bibinfo {author} {\bibfnamefont {D.}~\bibnamefont {Sanga}}, \bibinfo
  {author} {\bibfnamefont {M.}~\bibnamefont {Sano}}, \ and\ \bibinfo {author}
  {\bibfnamefont {T.}~\bibnamefont {Mukai}},\ }\href {\doibase
  10.1088/0022-3727/43/35/354002} {\bibfield  {journal} {\bibinfo  {journal}
  {J. Phys. D: Appl. Phys.}\ }\textbf {\bibinfo {volume} {43}},\ \bibinfo
  {pages} {354002} (\bibinfo {year} {2010})}\BibitemShut {NoStop}%
\bibitem [{\citenamefont {Aleksandrov}\ \emph {et~al.}(2000)\citenamefont
  {Aleksandrov}, \citenamefont {Gavrikova},\ and\ \citenamefont
  {Zykov}}]{Aleksandrov2000}%
  \BibitemOpen
  \bibfield  {author} {\bibinfo {author} {\bibfnamefont {S.~E.}\ \bibnamefont
  {Aleksandrov}}, \bibinfo {author} {\bibfnamefont {T.~A.}\ \bibnamefont
  {Gavrikova}}, \ and\ \bibinfo {author} {\bibfnamefont {V.~A.}\ \bibnamefont
  {Zykov}},\ }\href {\doibase 10.1134/1.1187973} {\bibfield  {journal}
  {\bibinfo  {journal} {Semiconductors}\ }\textbf {\bibinfo {volume} {34}},\
  \bibinfo {pages} {291} (\bibinfo {year} {2000})}\BibitemShut {NoStop}%
\bibitem [{\citenamefont {Ozgit}\ \emph {et~al.}(2012)\citenamefont
  {Ozgit}, \citenamefont {Donmez}, \citenamefont {Alevi},\ and\ \citenamefont
  {Biyikli}}]{Ozgit2012}%
  \BibitemOpen
  \bibfield  {author} {\bibinfo {author} {\bibfnamefont {C.}\ \bibnamefont
  {Ozgit}}, \bibinfo {author} {\bibfnamefont {I.}\ \bibnamefont
  {Donmez}}, \bibinfo {author} {\bibfnamefont {M.}\ \bibnamefont
  {Alevli}}, \ and\ \bibinfo {author} {\bibfnamefont {N.}\ \bibnamefont
  {Biyikli}},\ }\href {\doibase 10.1116/1.3664102} {\bibfield  {journal}
  {\bibinfo  {journal} {J. Vac. Sci. Technol. A}\ }\textbf {\bibinfo {volume} {30}},\
  \bibinfo {pages} {01A124} (\bibinfo {year} {2012})}\BibitemShut {NoStop}%
 \bibitem [{\citenamefont {Obinata}\ \emph {et~al.}(2000)\citenamefont
  {Obinata}, \citenamefont {Sugimoto},\ \citenamefont {Ijima},\ 
  \citenamefont {Ishibiki},\ \citenamefont {Egawa},\ \citenamefont {Honda},
  \ and\ \citenamefont  {Kawanashi}}]{Obinata2005}%
  \BibitemOpen
  \bibfield  {author} {\bibinfo {author} {\bibfnamefont {N.}\ \bibnamefont
  {Obinata}},\  \bibinfo {author} {\bibfnamefont {K.}\ \bibnamefont
  {Sugimoto}},\ \bibinfo {author} {\bibfnamefont {K.}\ \bibnamefont
  {Ishibiki}}, \bibinfo {author} {\bibfnamefont {S.}\ \bibnamefont
  {Egawa}}, \bibinfo {author} {\bibfnamefont {T.}\ \bibnamefont
  {Honda}},
   \ and\ \bibinfo {author} {\bibfnamefont {H.}\ \bibnamefont
  {Kawanishi}},\ }\href {\doibase 10.1143/JJAP.44.8432} {\bibfield  {journal}
  {\bibinfo  {journal} {Jpn. J. Appl. Phys.}\ }\textbf {\bibinfo {volume} {44}},\
  \bibinfo {pages} {8432} (\bibinfo {year} {2005})}\BibitemShut {NoStop}%
\bibitem [{\citenamefont {Sawada}\ \emph {et~al.}(2007)\citenamefont
  {Sawada}, \citenamefont {Sawadaishi},\ \citenamefont {Yamamoto},\ 
  \citenamefont {Arai},\ and\ \citenamefont  {Honda}}]{Sawada2007}%
  \BibitemOpen
  \bibfield  {author} {\bibinfo {author} {\bibfnamefont {M.}\ \bibnamefont
  {Sawada}}, \bibinfo {author} {\bibfnamefont {M.}\ \bibnamefont
  {Sawadaishi}},\ \bibinfo {author} {\bibfnamefont {H.}\ \bibnamefont
  {Yamamoto}}, \bibinfo {author} {\bibfnamefont {M.}\ \bibnamefont
  {Arai}},
   \ and\ \bibinfo {author} {\bibfnamefont {T.}\ \bibnamefont
  {Honda}},\ }\href {\doibase 10.1016/j.jcrysgro.2006.11.062} {\bibfield  {journal}
  {\bibinfo  {journal} {J. Cryst. Growth}\ }\textbf {\bibinfo {volume} {301-302}},\
  \bibinfo {pages} {67} (\bibinfo {year} {2007})}\BibitemShut {NoStop}%
%
\bibitem [{\citenamefont {Fern\'{a}ndez-Garrido}\ \emph {et~al.}(2007)\citenamefont
  {Fern\'{a}ndez-Garrido}, \citenamefont {Koblm\"{u}ller},\ \citenamefont {Calleja},\ and\ \citenamefont  {Speck}}]{Fernandez-Garrido2008}%
  \BibitemOpen
  \bibfield  {author} {\bibinfo {author} {\bibfnamefont {S.}\ \bibnamefont
  {Fern\'{a}ndez-Garrido}}, \bibinfo {author} {\bibfnamefont {G.}\ \bibnamefont
  {Koblm\"{u}ller}},\ \bibinfo {author} {\bibfnamefont {E.}\ \bibnamefont
  {Calleja}},\ and\ \bibinfo {author} {\bibfnamefont {J.S.}\ \bibnamefont
  {Speck}},\ }\href {\doibase 10.1063/1.2968442} {\bibfield  {journal}
  {\bibinfo  {journal} {J. Appl. Phys.}\ }\textbf {\bibinfo {volume} {104}},\
  \bibinfo {pages} {033541} (\bibinfo {year} {2008})}\BibitemShut {NoStop}%
%
\bibitem [{\citenamefont {Zapol}\ \emph {et~al.}(1997)\citenamefont {Zapol},
  \citenamefont {Pandey},\ and\ \citenamefont {Gale}}]{Zapol1997}%
  \BibitemOpen
  \bibfield  {author} {\bibinfo {author} {\bibfnamefont {P.}~\bibnamefont
  {Zapol}}, \bibinfo {author} {\bibfnamefont {R.}~\bibnamefont {Pandey}}, \
  and\ \bibinfo {author} {\bibfnamefont {J.~D.}\ \bibnamefont {Gale}},\ }\href
  {\doibase 10.1088/0953-8984/9/44/008} {\bibfield  {journal} {\bibinfo
  {journal} {J. Phys.: Condens. Matter}\ }\textbf {\bibinfo {volume} {9}},\
  \bibinfo {pages} {9517} (\bibinfo {year} {1997})}\BibitemShut {NoStop}%
\bibitem [{\citenamefont {Catlow}\ \emph {et~al.}(2010)\citenamefont {Catlow},
  \citenamefont {Guo}, \citenamefont {Miskufova}, \citenamefont {Shevlin},
  \citenamefont {Smith}, \citenamefont {Sokol}, \citenamefont {Walsh},
  \citenamefont {Wilson},\ and\ \citenamefont {Woodley}}]{Catlow2010}%
  \BibitemOpen
  \bibfield  {author} {\bibinfo {author} {\bibfnamefont {C.~R.~A.}\
  \bibnamefont {Catlow}}, \bibinfo {author} {\bibfnamefont {Z.~X.}\
  \bibnamefont {Guo}}, \bibinfo {author} {\bibfnamefont {M.}~\bibnamefont
  {Miskufova}}, \bibinfo {author} {\bibfnamefont {S.~A.}\ \bibnamefont
  {Shevlin}}, \bibinfo {author} {\bibfnamefont {A.~G.~H.}\ \bibnamefont
  {Smith}}, \bibinfo {author} {\bibfnamefont {A.~A.}\ \bibnamefont {Sokol}},
  \bibinfo {author} {\bibfnamefont {A.}~\bibnamefont {Walsh}}, \bibinfo
  {author} {\bibfnamefont {D.~J.}\ \bibnamefont {Wilson}}, \ and\ \bibinfo
  {author} {\bibfnamefont {S.~M.}\ \bibnamefont {Woodley}},\ }\href {\doibase
  10.1098/rsta.2010.0111} {\bibfield  {journal} {\bibinfo  {journal}
  {Phil. Trans. Roy. Soc. A}\ }\textbf {\bibinfo {volume} {368}},\ \bibinfo {pages}
  {3379} (\bibinfo {year} {2010})}\BibitemShut {NoStop}%
\bibitem [{\citenamefont {Lany}\ and\ \citenamefont {Zunger}(2010)}]{Lany2010}%
  \BibitemOpen
  \bibfield  {author} {\bibinfo {author} {\bibfnamefont {S.}~\bibnamefont
  {Lany}}\ and\ \bibinfo {author} {\bibfnamefont {A.}~\bibnamefont {Zunger}},\
  }\href {\doibase 10.1063/1.3383236} {\bibfield  {journal} {\bibinfo
  {journal} {Appl. Phys. Lett.}\ }\textbf {\bibinfo {volume} {96}},\ \bibinfo
  {pages} {142114} (\bibinfo {year} {2010})}\BibitemShut {NoStop}%
\bibitem [{\citenamefont {Lambrecht}\ \emph {et~al.}(1994)\citenamefont
  {Lambrecht}, \citenamefont {Segall}, \citenamefont {Strite},\ and\
  \citenamefont {Martin}}]{Lambrecht1994}%
  \BibitemOpen
  \bibfield  {author} {\bibinfo {author} {\bibfnamefont {W.}~\bibnamefont
  {Lambrecht}}, \bibinfo {author} {\bibfnamefont {B.}~\bibnamefont {Segall}},
  \bibinfo {author} {\bibfnamefont {S.}~\bibnamefont {Strite}}, \ and\ \bibinfo
  {author} {\bibfnamefont {G.}~\bibnamefont {Martin}},\ }\href
  {http://prb.aps.org/abstract/PRB/v50/i19/p14155\_1} {\bibfield  {journal}
  {\bibinfo  {journal} {Phys. Rev. B}\ }\textbf {\bibinfo {volume} {50}},\
  \bibinfo {pages} {155} (\bibinfo {year} {1994})}\BibitemShut {NoStop}%
\bibitem [{\citenamefont {Limpijumnong}\ \emph {et~al.}(2003)\citenamefont
  {Limpijumnong}, \citenamefont {Northrup},\ and\ \citenamefont {{Van de
  Walle}}}]{Limpijumnong2003}%
  \BibitemOpen
  \bibfield  {author} {\bibinfo {author} {\bibfnamefont {S.}~\bibnamefont
  {Limpijumnong}}, \bibinfo {author} {\bibfnamefont {J.}~\bibnamefont
  {Northrup}}, \ and\ \bibinfo {author} {\bibfnamefont {C.}~\bibnamefont {{Van
  de Walle}}},\ }\href {\doibase 10.1103/PhysRevB.68.075206} {\bibfield
  {journal} {\bibinfo  {journal} {Phys. Rev. B}\ }\textbf {\bibinfo {volume}
  {68}},\ \bibinfo {pages} {075206} (\bibinfo {year} {2003})}\BibitemShut
  {NoStop}%
  \bibitem [{\citenamefont {Carter}\ \emph{et~al.}(2009)}]{Carter2008}%
  \BibitemOpen
  \bibfield  {author} {\bibinfo {author} {\bibfnamefont {D.~J.}~\bibnamefont
  {Carter}}, \bibinfo {author} {\bibfnamefont {J.~D.}~\bibnamefont
  {Gale}}, \bibinfo {author} {\bibfnamefont {B.}~\bibnamefont {Delley}}, and \ \bibinfo {author} {\bibnamefont {C.}~\bibnamefont {Stampfl}},\ }\href {\doibase 10.1103/PhysRevB.77.115349} {\bibfield
  {journal} {\bibinfo  {journal} {Phys. Rev. B}\ }\textbf {\bibinfo
  {volume} {77}},\ \bibinfo {pages} {115349} (\bibinfo {year}
  {2008})}\BibitemShut {NoStop}%
    \bibitem [{\citenamefont {Neugebauer}\ and\ \citenamefont {{Van de
  Walle}}(1994)}]{Neugebauer1994}%
  \BibitemOpen
  \bibfield  {author} {\bibinfo {author} {\bibfnamefont {J.}~\bibnamefont
  {Neugebauer}}\ and\ \bibinfo {author} {\bibfnamefont {C.~G.}\ \bibnamefont
  {{Van de Walle}}},\ }\href {\doibase 10.1103/PhysRevB.50.8067} {\bibfield
  {journal} {\bibinfo  {journal} {Phys. Rev. B}\ }\textbf {\bibinfo {volume}
  {50}},\ \bibinfo {pages} {8067} (\bibinfo {year} {1994})}\BibitemShut
  {NoStop}%
\bibitem [{\citenamefont {Zywietz}\ \emph {et~al.}(1999)\citenamefont
  {Zywietz}, \citenamefont {Neugebauer},\ and\ \citenamefont
  {Scheffler}}]{Zywietz1999}%
  \BibitemOpen
  \bibfield  {author} {\bibinfo {author} {\bibfnamefont {T.~K.}\ \bibnamefont
  {Zywietz}}, \bibinfo {author} {\bibfnamefont {J.}~\bibnamefont {Neugebauer}},
  \ and\ \bibinfo {author} {\bibfnamefont {M.}~\bibnamefont {Scheffler}},\
  }\href {\doibase 10.1063/1.123658} {\bibfield  {journal} {\bibinfo  {journal}
  {Appl. Phys. Lett.}\ }\textbf {\bibinfo {volume} {74}},\ \bibinfo {pages}
  {1695} (\bibinfo {year} {1999})}\BibitemShut {NoStop}%
\bibitem [{\citenamefont {Hohenberg}\ and\ \citenamefont
  {Kohn}(1964)}]{Hohenberg1964}%
  \BibitemOpen
  \bibfield  {author} {\bibinfo {author} {\bibfnamefont {P.}~\bibnamefont
  {Hohenberg}}\ and\ \bibinfo {author} {\bibfnamefont {W.}~\bibnamefont
  {Kohn}},\ }\href {\doibase 10.1103/PhysRev.136.B864} {\bibfield  {journal}
  {\bibinfo  {journal} {Phys. Rev. B}\ }\textbf {\bibinfo {volume} {136}},\
  \bibinfo {pages} {864} (\bibinfo {year} {1964})}\BibitemShut {NoStop}%
\bibitem [{\citenamefont {Kohn}\ and\ \citenamefont {Sham}(1965)}]{Kohn1965}%
  \BibitemOpen
  \bibfield  {author} {\bibinfo {author} {\bibfnamefont {W.}~\bibnamefont
  {Kohn}}\ and\ \bibinfo {author} {\bibfnamefont {L.}~\bibnamefont {Sham}},\
  }\href {http://link.aps.org/doi/10.1103/PhysRev.140.A1133} {\bibfield
  {journal} {\bibinfo  {journal} {Phys. Rev. A}\ }\textbf {\bibinfo {volume}
  {140}},\ \bibinfo {pages} {1133} (\bibinfo {year} {1965})}\BibitemShut
  {NoStop}%
\bibitem [{\citenamefont {Stoffel}\ \emph {et~al.}(2010)\citenamefont
  {Stoffel}, \citenamefont {Wessel}, \citenamefont {Lumey},\ and\ \citenamefont
  {Dronskowski}}]{Stoffel2010}%
  \BibitemOpen
  \bibfield  {author} {\bibinfo {author} {\bibfnamefont {R.~P.}\ \bibnamefont
  {Stoffel}}, \bibinfo {author} {\bibfnamefont {C.}~\bibnamefont {Wessel}},
  \bibinfo {author} {\bibfnamefont {M.-W.}\ \bibnamefont {Lumey}}, \ and\
  \bibinfo {author} {\bibfnamefont {R.}~\bibnamefont {Dronskowski}},\ }\href
  {\doibase 10.1002/anie.200906780} {\bibfield  {journal} {\bibinfo  {journal}
  {Angew. Chem. Int. Ed.}\ }\textbf {\bibinfo
  {volume} {49}},\ \bibinfo {pages} {5242} (\bibinfo {year}
  {2010})}\BibitemShut {NoStop}%
\bibitem [{\citenamefont {Blum}\ \emph {et~al.}(2009)\citenamefont {Blum},
  \citenamefont {Gehrke}, \citenamefont {Hanke}, \citenamefont {Havu},
  \citenamefont {Havu}, \citenamefont {Ren}, \citenamefont {Reuter},\ and\
  \citenamefont {Scheffler}}]{Blum2009}%
  \BibitemOpen
  \bibfield  {author} {\bibinfo {author} {\bibfnamefont {V.}~\bibnamefont
  {Blum}}, \bibinfo {author} {\bibfnamefont {R.}~\bibnamefont {Gehrke}},
  \bibinfo {author} {\bibfnamefont {F.}~\bibnamefont {Hanke}}, \bibinfo
  {author} {\bibfnamefont {P.}~\bibnamefont {Havu}}, \bibinfo {author}
  {\bibfnamefont {V.}~\bibnamefont {Havu}}, \bibinfo {author} {\bibfnamefont
  {X.}~\bibnamefont {Ren}}, \bibinfo {author} {\bibfnamefont {K.}~\bibnamefont
  {Reuter}}, \ and\ \bibinfo {author} {\bibfnamefont {M.}~\bibnamefont
  {Scheffler}},\ }\href {\doibase 10.1016/j.cpc.2009.06.022} {\bibfield
  {journal} {\bibinfo  {journal} {Comput. Phys. Commun.}\ }\textbf {\bibinfo
  {volume} {180}},\ \bibinfo {pages} {2175} (\bibinfo {year}
  {2009})}\BibitemShut {NoStop}%
\bibitem [{\citenamefont {Havu}\ \emph {et~al.}(2009)\citenamefont {Havu},
  \citenamefont {Blum}, \citenamefont {Havu},\ and\ \citenamefont
  {Scheffler}}]{Havu2009}%
  \BibitemOpen
  \bibfield  {author} {\bibinfo {author} {\bibfnamefont {V.}~\bibnamefont
  {Havu}}, \bibinfo {author} {\bibfnamefont {V.}~\bibnamefont {Blum}}, \bibinfo
  {author} {\bibfnamefont {P.}~\bibnamefont {Havu}}, \ and\ \bibinfo {author}
  {\bibfnamefont {M.}~\bibnamefont {Scheffler}},\ }\href {\doibase
  10.1016/j.jcp.2009.08.008} {\bibfield  {journal} {\bibinfo  {journal} {J.
  Comput. Phys.}\ }\textbf {\bibinfo {volume} {228}},\ \bibinfo {pages} {8367}
  (\bibinfo {year} {2009})}\BibitemShut {NoStop}%
\bibitem [{\citenamefont {Perdew}\ \emph {et~al.}(2008)\citenamefont {Perdew},
  \citenamefont {Ruzsinszky}, \citenamefont {Csonka}, \citenamefont {Vydrov},
  \citenamefont {Scuseria}, \citenamefont {Constantin}, \citenamefont {Zhou},\
  and\ \citenamefont {Burke}}]{Perdew2008}%
  \BibitemOpen
  \bibfield  {author} {\bibinfo {author} {\bibfnamefont {J.}~\bibnamefont
  {Perdew}}, \bibinfo {author} {\bibfnamefont {A.}~\bibnamefont {Ruzsinszky}},
  \bibinfo {author} {\bibfnamefont {G.}~\bibnamefont {Csonka}}, \bibinfo
  {author} {\bibfnamefont {O.}~\bibnamefont {Vydrov}}, \bibinfo {author}
  {\bibfnamefont {G.}~\bibnamefont {Scuseria}}, \bibinfo {author}
  {\bibfnamefont {L.}~\bibnamefont {Constantin}}, \bibinfo {author}
  {\bibfnamefont {X.}~\bibnamefont {Zhou}}, \ and\ \bibinfo {author}
  {\bibfnamefont {K.}~\bibnamefont {Burke}},\ }\href {\doibase
  10.1103/PhysRevLett.100.136406} {\bibfield  {journal} {\bibinfo  {journal}
  {Phys. Rev. Lett.}\ }\textbf {\bibinfo {volume} {100}},\ \bibinfo {pages}
  {136406} (\bibinfo {year} {2008})}\BibitemShut {NoStop}%
\bibitem [{\citenamefont {Csonka}\ \emph {et~al.}(2009)\citenamefont {Csonka},
  \citenamefont {Perdew}, \citenamefont {Ruzsinszky}, \citenamefont
  {Philipsen}, \citenamefont {Leb\`{e}gue}, \citenamefont {Paier},
  \citenamefont {Vydrov},\ and\ \citenamefont {\'{A}ngy\'{a}n}}]{Csonka2009}%
  \BibitemOpen
  \bibfield  {author} {\bibinfo {author} {\bibfnamefont {G.}~\bibnamefont
  {Csonka}}, \bibinfo {author} {\bibfnamefont {J.}~\bibnamefont {Perdew}},
  \bibinfo {author} {\bibfnamefont {A.}~\bibnamefont {Ruzsinszky}}, \bibinfo
  {author} {\bibfnamefont {P.}~\bibnamefont {Philipsen}}, \bibinfo {author}
  {\bibfnamefont {S.}~\bibnamefont {Leb\`{e}gue}}, \bibinfo {author}
  {\bibfnamefont {J.}~\bibnamefont {Paier}}, \bibinfo {author} {\bibfnamefont
  {O.}~\bibnamefont {Vydrov}}, \ and\ \bibinfo {author} {\bibfnamefont
  {J.}~\bibnamefont {\'{A}ngy\'{a}n}},\ }\href {\doibase
  10.1103/PhysRevB.79.155107} {\bibfield  {journal} {\bibinfo  {journal} {Phys.
  Rev. B}\ }\textbf {\bibinfo {volume} {79}},\ \bibinfo {pages} {155107}
  (\bibinfo {year} {2009})}\BibitemShut {NoStop}%
\bibitem [{\citenamefont {Lany}(2008)}]{Lany2008}%
  \BibitemOpen
  \bibfield  {author} {\bibinfo {author} {\bibfnamefont {S.}~\bibnamefont
  {Lany}},\ }\href {\doibase 10.1103/PhysRevB.78.245207} {\bibfield  {journal}
  {\bibinfo  {journal} {Phys. Rev. B}\ }\textbf {\bibinfo {volume} {78}},\
  \bibinfo {pages} {1} (\bibinfo {year} {2008})}\BibitemShut {NoStop}%
\bibitem [{\citenamefont {Xiao}\ \emph {et~al.}(2011)\citenamefont {Xiao},
  \citenamefont {Tahir-Kheli},\ and\ \citenamefont {Goddard}}]{Xiao2011}%
  \BibitemOpen
  \bibfield  {author} {\bibinfo {author} {\bibfnamefont {H.}~\bibnamefont
  {Xiao}}, \bibinfo {author} {\bibfnamefont {J.}~\bibnamefont {Tahir-Kheli}}, \
  and\ \bibinfo {author} {\bibfnamefont {W.~A.}\ \bibnamefont {Goddard}},\
  }\href {\doibase 10.1021/jz101565j} {\bibfield  {journal} {\bibinfo
  {journal} {J. Phys. Chem. Lett.}\ }\textbf {\bibinfo
  {volume} {2}},\ \bibinfo {pages} {212} (\bibinfo {year} {2011})}\BibitemShut
  {NoStop}%
\bibitem [{\citenamefont {Paszkowicz}\ \emph {et~al.}(2004)\citenamefont
  {Paszkowicz}, \citenamefont {Podsiad\l{}o},\ and\ \citenamefont
  {Minikayev}}]{Paszkowicz2004}%
  \BibitemOpen
  \bibfield  {author} {\bibinfo {author} {\bibfnamefont {W.}~\bibnamefont
  {Paszkowicz}}, \bibinfo {author} {\bibfnamefont {S.}~\bibnamefont
  {Podsiad\l{}o}}, \ and\ \bibinfo {author} {\bibfnamefont {R.}~\bibnamefont
  {Minikayev}},\ }\href {\doibase 10.1016/j.jallcom.2004.05.036} {\bibfield
  {journal} {\bibinfo  {journal} {J. Alloys Compd.}\ }\textbf {\bibinfo
  {volume} {382}},\ \bibinfo {pages} {100} (\bibinfo {year}
  {2004})}\BibitemShut {NoStop}%
\bibitem [{\citenamefont {\AA{}hman}\ \emph {et~al.}(1996)\citenamefont
  {\AA~hman}, \citenamefont {Svensson},\ and\ \citenamefont
  {Albertsson}}]{Ahman1996}%
  \BibitemOpen
  \bibfield  {author} {\bibinfo {author} {\bibfnamefont {J.}~\bibnamefont
  {\AA~hman}}, \bibinfo {author} {\bibfnamefont {G.}~\bibnamefont {Svensson}},
  \ and\ \bibinfo {author} {\bibfnamefont {J.}~\bibnamefont {Albertsson}},\
  }\href {\doibase 10.1107/S0108270195016404} {\bibfield  {journal} {\bibinfo
  {journal} {Acta Crystallogr. C}\ }\textbf {\bibinfo {volume} {52}},\ \bibinfo {pages} {1336}
  (\bibinfo {year} {1996})}\BibitemShut {NoStop}%
\bibitem [{\citenamefont {Bradley}(1935)}]{Bradley1935}%
  \BibitemOpen
  \bibfield  {author} {\bibinfo {author} {\bibfnamefont {A.}~\bibnamefont
  {Bradley}},\ }\href@noop {} {\bibfield  {journal} {\bibinfo  {journal}
  {Z. Kristallogr. Kristallgeom. Kristallphys. Kristallchem.}\ }\textbf {\bibinfo {volume} {91}},\ \bibinfo {pages} {302}
  (\bibinfo {year} {1935})}\BibitemShut {NoStop}%
\bibitem [{\citenamefont {Moreno}\ and\ \citenamefont
  {Soler}(1992)}]{Moreno1992}%
  \BibitemOpen
  \bibfield  {author} {\bibinfo {author} {\bibfnamefont {J.}~\bibnamefont
  {Moreno}}\ and\ \bibinfo {author} {\bibfnamefont {J.}~\bibnamefont {Soler}},\
  }\href {\doibase 10.1103/PhysRevB.45.13891} {\bibfield  {journal} {\bibinfo
  {journal} {Phys. Rev. B}\ }\textbf {\bibinfo {volume} {45}},\ \bibinfo
  {pages} {13891} (\bibinfo {year} {1992})}\BibitemShut {NoStop}%
\bibitem [{\citenamefont {Persson}\ \emph {et~al.}(2005)\citenamefont
  {Persson}, \citenamefont {Zhao}, \citenamefont {Lany},\ and\ \citenamefont
  {Zunger}}]{Persson2005}%
  \BibitemOpen
  \bibfield  {author} {\bibinfo {author} {\bibfnamefont {C.}~\bibnamefont
  {Persson}}, \bibinfo {author} {\bibfnamefont {Y.-J.}\ \bibnamefont {Zhao}},
  \bibinfo {author} {\bibfnamefont {S.}~\bibnamefont {Lany}}, \ and\ \bibinfo
  {author} {\bibfnamefont {A.}~\bibnamefont {Zunger}},\ }\href {\doibase
  10.1103/PhysRevB.72.035211} {\bibfield  {journal} {\bibinfo  {journal} {Phys.
  Rev. B}\ }\textbf {\bibinfo {volume} {72}},\ \bibinfo {pages} {035211}
  (\bibinfo {year} {2005})}\BibitemShut {NoStop}%
\bibitem [{\citenamefont {Lany}\ and\ \citenamefont
  {Zunger}(2008)}]{Lany2008a}%
  \BibitemOpen
  \bibfield  {author} {\bibinfo {author} {\bibfnamefont {S.}~\bibnamefont
  {Lany}}\ and\ \bibinfo {author} {\bibfnamefont {A.}~\bibnamefont {Zunger}},\
  }\href {\doibase 10.1103/PhysRevB.78.235104} {\bibfield  {journal} {\bibinfo
  {journal} {Phys. Rev. B}\ }\textbf {\bibinfo {volume} {78}},\ \bibinfo
  {pages} {235104} (\bibinfo {year} {2008})}\BibitemShut {NoStop}%
\bibitem [{\citenamefont {Irikura}(2007)}]{Irikura2007}%
  \BibitemOpen
  \bibfield  {author} {\bibinfo {author} {\bibfnamefont {K.~K.}\ \bibnamefont
  {Irikura}},\ }\href {\doibase 10.1063/1.2436891} {\bibfield  {journal}
  {\bibinfo  {journal} {J. Phys. Chem. Ref. Data}\ }\textbf {\bibinfo {volume}
  {36}},\ \bibinfo {pages} {389} (\bibinfo {year} {2007})}\BibitemShut
  {NoStop}%
\bibitem [{\citenamefont {Haynes}\ and\ \citenamefont
  {Lide}(2011)}]{handbook-crc}%
  \BibitemOpen
  \bibfield  {author} {\bibinfo {author} {\bibfnamefont {W.~M.}\ \bibnamefont
  {Haynes}}\ and\ \bibinfo {author} {\bibfnamefont {D.~R.}\ \bibnamefont
  {Lide}},\ }\href {http://books.google.co.uk/books?id=N7-8cQAACAAJ} {\emph
  {\bibinfo {title} {{CRC Handbook of Chemistry and Physics}}}},\  (\bibinfo  {publisher} {Taylor \& Francis},\
  \bibinfo {year} {2011})\BibitemShut {NoStop}%
\bibitem [{\citenamefont {Reuter}\ \emph {et~al.}(2005)\citenamefont {Reuter},
  \citenamefont {Stampfl},\ and\ \citenamefont {Scheffler}}]{Reuter2005}%
  \BibitemOpen
  \bibfield  {author} {\bibinfo {author} {\bibfnamefont {K.}~\bibnamefont
  {Reuter}}, \bibinfo {author} {\bibfnamefont {C.}~\bibnamefont {Stampfl}}, \
  and\ \bibinfo {author} {\bibfnamefont {M.}~\bibnamefont {Scheffler}},\
  }\href@noop {} {\emph {\bibinfo {title} {{Handbook of Materials Modeling,
  Part A. Methods}}}},\ edited by\ \bibinfo {editor} {\bibfnamefont
  {S.}~\bibnamefont {Yip}}\ (\bibinfo  {publisher} {Springer},\ \bibinfo
  {address} {Berlin},\ \bibinfo {year} {2005})\ pp.\ \bibinfo {pages}
  {149--234}\BibitemShut {NoStop}%
\bibitem [{\citenamefont {Reuter}\ and\ \citenamefont
  {Scheffler}(2001)}]{Reuter2001}%
  \BibitemOpen
  \bibfield  {author} {\bibinfo {author} {\bibfnamefont {K.}~\bibnamefont
  {Reuter}}\ and\ \bibinfo {author} {\bibfnamefont {M.}~\bibnamefont
  {Scheffler}},\ }\href {\doibase 10.1103/PhysRevB.65.035406} {\bibfield
  {journal} {\bibinfo  {journal} {Phys. Rev. B}\ }\textbf {\bibinfo {volume}
  {65}},\ \bibinfo {pages} {035406} (\bibinfo {year} {2001})}\BibitemShut
  {NoStop}%
\bibitem [{\citenamefont {Morgan}\ and\ \citenamefont
  {Watson}(2010)}]{Morgan2010}%
  \BibitemOpen
  \bibfield  {author} {\bibinfo {author} {\bibfnamefont {B.~J.}\ \bibnamefont
  {Morgan}}\ and\ \bibinfo {author} {\bibfnamefont {G.~W.}\ \bibnamefont
  {Watson}},\ }\href {\doibase 10.1021/jp9088047} {\bibfield  {journal}
  {\bibinfo  {journal} {J. Phys. Chem. C}\ }\textbf
  {\bibinfo {volume} {114}},\ \bibinfo {pages} {2321} (\bibinfo {year}
  {2010})}\BibitemShut {NoStop}%
\bibitem [{\citenamefont {Togo}\ \emph {et~al.}(2008)\citenamefont {Togo},
  \citenamefont {Oba},\ and\ \citenamefont {Tanaka}}]{Togo2008}%
  \BibitemOpen
  \bibfield  {author} {\bibinfo {author} {\bibfnamefont {A.}~\bibnamefont
  {Togo}}, \bibinfo {author} {\bibfnamefont {F.}~\bibnamefont {Oba}}, \ and\
  \bibinfo {author} {\bibfnamefont {I.}~\bibnamefont {Tanaka}},\ }\href
  {\doibase 10.1103/PhysRevB.78.134106} {\bibfield  {journal} {\bibinfo
  {journal} {Phys. Rev. B}\ }\textbf {\bibinfo {volume} {78}},\ \bibinfo
  {pages} {134106} (\bibinfo {year} {2008})}\BibitemShut {NoStop}%
  \bibitem [{\citenamefont {Parlinski},\ \citenamefont {Li},\ and\ \citenamefont {Kawazoe}(1997)}]{Parlinski1997}
	\bibitemOpen
	\bibfield  {author} {\bibinfo {author} {\bibnamefont {K. }~\bibnamefont{Parlinski}},\ \bibinfo {author} {\bibnamefont {Z.~Q.}~\bibnamefont {Li}},\ and \bibinfo {author} {\bibnamefont {Y.}~\bibnamefont{Kawazoe}},\ }\href {\doibase 10.1103/PhysRevLett.78.4063} {\bibfield {journal} {\bibinfo {journal} {Phys. Rev. Lett.}\ }\textbf {\bibinfo {volume} {78}},\ \bibinfo {pages} {4063} (\bibinfo {year} {1997})}\BibitemShut {NoStop}
\bibitem [{\citenamefont {Sarasamak}\ \emph {et~al.}(2010)\citenamefont
  {Sarasamak}, \citenamefont {Limpijumnong},\ and\ \citenamefont
  {Lambrecht}}]{Sarasamak2010}%
  \BibitemOpen
  \bibfield  {author} {\bibinfo {author} {\bibfnamefont {K.}~\bibnamefont
  {Sarasamak}}, \bibinfo {author} {\bibfnamefont {S.}~\bibnamefont
  {Limpijumnong}}, \ and\ \bibinfo {author} {\bibfnamefont {W.~R.~L.}\
  \bibnamefont {Lambrecht}},\ }\href {\doibase 10.1103/PhysRevB.82.035201}
  {\bibfield  {journal} {\bibinfo  {journal} {Phys. Rev. B}\ }\textbf {\bibinfo
  {volume} {82}},\ \bibinfo {pages} {035201} (\bibinfo {year}
  {2010})}\BibitemShut {NoStop}%
\bibitem [{\citenamefont {Chase}(1998)}]{Chase1998}%
  \BibitemOpen
  \bibfield  {author} {\bibinfo {author} {\bibfnamefont {M.~W.~J.}\ \bibnamefont
  {Chase}},\ }\href@noop {} {\emph {\bibinfo {title} {{NIST-JANAF Thermochemical Tables}}}}\ \bibinfo {publisher} {American Chemical Society},\ \bibinfo {address} {New York},\ \bibinfo {year} {1998})\BibitemShut {NoStop}%
\bibitem [{\citenamefont {Linstrom}\ and\ \citenamefont
  {Mallard}(2005)}]{Linstrom2005}%
  \BibitemOpen
  \bibfield  {author} {\bibinfo {author} {\bibfnamefont {P.}~\bibnamefont
  {Linstrom}}\ and\ \bibinfo {author} {\bibfnamefont {W.}~\bibnamefont
  {Mallard}},\ }\href@noop {} {\emph {\bibinfo {title} {{NIST Chemistry
  WebBook, Nist Standard Reference database Number 69}}}}\ (\bibinfo
  {publisher} {National Institute of Standards and Technology},\ \bibinfo
  {address} {Gaithersburg MD},\ \bibinfo {year} {2005})\BibitemShut {NoStop}%
\bibitem [{\citenamefont {Shomate}(1944)}]{Shomate1944}%
  \BibitemOpen
  \bibfield  {author} {\bibinfo {author} {\bibfnamefont {C.}~\bibnamefont
  {Shomate}},\ }\href {\doibase 10.1021/ja01234a025} {\bibfield  {journal}
  {\bibinfo  {journal} {J. Am. Chem. Soc.}\ }\textbf {\bibinfo {volume} {66}},\
  \bibinfo {pages} {928} (\bibinfo {year} {1944})}\BibitemShut {NoStop}%
\bibitem [{\citenamefont {Shomate}(1954)}]{Shomate1954}%
  \BibitemOpen
  \bibfield  {author} {\bibinfo {author} {\bibfnamefont {C.}~\bibnamefont
  {Shomate}},\ }\href {\doibase 10.1021/j150514a018} {\bibfield  {journal}
  {\bibinfo  {journal} {J. Phys. Chem.}\ }\textbf {\bibinfo
  {volume} {58}},\ \bibinfo {pages} {368} (\bibinfo {year} {1954})}\BibitemShut
  {NoStop}%
\bibitem [{\citenamefont {Warn}\ and\ \citenamefont {Peters}(1996)}]{Warn1996}%
  \BibitemOpen
  \bibfield  {author} {\bibinfo {author} {\bibfnamefont {J.}~\bibnamefont
  {Warn}}\ and\ \bibinfo {author} {\bibfnamefont {A.}~\bibnamefont {Peters}},\
  }\href@noop {} {\emph {\bibinfo {title} {{Concise Chemical
  Thermodynamics}}}},\ \bibinfo {edition} {2nd}\ ed.\ (\bibinfo  {publisher}
  {Chapman \& Hall},\ \bibinfo {address} {London},\ \bibinfo {year} {1996})\
  pp.\ \bibinfo {pages} {92--93}\BibitemShut {NoStop}%
\bibitem [{\citenamefont {Danilchenko}\ \emph {et~al.}(2006)\citenamefont
  {Danilchenko}, \citenamefont {Paszkiewicz}, \citenamefont {Wolski},
  \citenamefont {Jezowski},\ and\ \citenamefont
  {Plackowski}}]{Danilchenko2006}%
  \BibitemOpen
  \bibfield  {author} {\bibinfo {author} {\bibfnamefont {B.~A.}\ \bibnamefont
  {Danilchenko}}, \bibinfo {author} {\bibfnamefont {T.}~\bibnamefont
  {Paszkiewicz}}, \bibinfo {author} {\bibfnamefont {S.}~\bibnamefont {Wolski}},
  \bibinfo {author} {\bibfnamefont {A.}~\bibnamefont {Jezowski}}, \ and\
  \bibinfo {author} {\bibfnamefont {T.}~\bibnamefont {Plackowski}},\ }\href
  {\doibase 10.1063/1.2335373} {\bibfield  {journal} {\bibinfo  {journal}
  {Appl. Phys. Lett.}\ }\textbf {\bibinfo {volume} {89}},\ \bibinfo {pages}
  {061901} (\bibinfo {year} {2006})}\BibitemShut {NoStop}%
\bibitem [{\citenamefont {Jacob}\ \emph {et~al.}(2007)\citenamefont {Jacob},
  \citenamefont {Singh},\ and\ \citenamefont {Waseda}}]{Jacob2007}%
  \BibitemOpen
  \bibfield  {author} {\bibinfo {author} {\bibfnamefont {K.}~\bibnamefont
  {Jacob}}, \bibinfo {author} {\bibfnamefont {S.}~\bibnamefont {Singh}}, \ and\
  \bibinfo {author} {\bibfnamefont {Y.}~\bibnamefont {Waseda}},\ }\href
  {\doibase 10.1557/JMR.2007.0441} {\bibfield  {journal} {\bibinfo  {journal}
  {J. Mater. Res.}\ }\textbf {\bibinfo {volume} {22}},\ \bibinfo {pages} {3475}
  (\bibinfo {year} {2007})}\BibitemShut {NoStop}%
\bibitem [{\citenamefont {Leitner}\ \emph {et~al.}(2003)\citenamefont
  {Leitner}, \citenamefont {Strejc}, \citenamefont {Sedmidubsk\'{y}},\ and\
  \citenamefont {R??\v{z}i\v{c}ka}}]{Leitner2003}%
  \BibitemOpen
  \bibfield  {author} {\bibinfo {author} {\bibfnamefont {J.}~\bibnamefont
  {Leitner}}, \bibinfo {author} {\bibfnamefont {A.}~\bibnamefont {Strejc}},
  \bibinfo {author} {\bibfnamefont {D.}~\bibnamefont {Sedmidubsk\'{y}}}, \ and\
  \bibinfo {author} {\bibfnamefont {K.}~\bibnamefont {R\r{u}\v{z}i\v{c}ka}},\
  }\href {\doibase 10.1016/S0040-6031(02)00547-6} {\bibfield  {journal}
  {\bibinfo  {journal} {Thermochim. Acta}\ }\textbf {\bibinfo {volume} {401}},\
  \bibinfo {pages} {169} (\bibinfo {year} {2003})}\BibitemShut {NoStop}%
\bibitem [{\citenamefont {{Adams, Jr.}}\ and\ \citenamefont
  {Johnston}(1952)}]{AdamsJr.1952}%
  \BibitemOpen
  \bibfield  {author} {\bibinfo {author} {\bibfnamefont {G.~B.}\ \bibnamefont
  {{Adams, Jr.}}}\ and\ \bibinfo {author} {\bibfnamefont {H.~L.}\ \bibnamefont
  {Johnston}},\ }\href {\doibase 10.1021/ja01139a018} {\bibfield  {journal}
  {\bibinfo  {journal} {J. Am. Chem. Soc.}\ }\textbf {\bibinfo {volume} {74}},\
  \bibinfo {pages} {4788} (\bibinfo {year} {1952})}\BibitemShut {NoStop}%
\bibitem [{\citenamefont {Mills}(1972)}]{Mills1972}%
  \BibitemOpen
  \bibfield  {author} {\bibinfo {author} {\bibfnamefont {K.}~\bibnamefont
  {Mills}},\ }\href@noop {} {\bibfield  {journal} {\bibinfo  {journal} {High
  Temp. -- High Press.}\ }\textbf {\bibinfo {volume} {4}},\ \bibinfo
  {pages} {371} (\bibinfo {year} {1972})}\BibitemShut {NoStop}%
\bibitem [{\citenamefont {Azuhata}\ \emph {et~al.}(1996)\citenamefont
  {Azuhata}, \citenamefont {Matsunaga}, \citenamefont {Shimada}, \citenamefont
  {Yoshida}, \citenamefont {Sota}, \citenamefont {Suzuki},\ and\ \citenamefont
  {Nakamura}}]{Azuhata1996}%
  \BibitemOpen
  \bibfield  {author} {\bibinfo {author} {\bibfnamefont {T.}~\bibnamefont
  {Azuhata}}, \bibinfo {author} {\bibfnamefont {T.}~\bibnamefont {Matsunaga}},
  \bibinfo {author} {\bibfnamefont {K.}~\bibnamefont {Shimada}}, \bibinfo
  {author} {\bibfnamefont {K.}~\bibnamefont {Yoshida}}, \bibinfo {author}
  {\bibfnamefont {T.}~\bibnamefont {Sota}}, \bibinfo {author} {\bibfnamefont
  {K.}~\bibnamefont {Suzuki}}, \ and\ \bibinfo {author} {\bibfnamefont
  {S.}~\bibnamefont {Nakamura}},\ }\href {\doibase
  10.1016/0921-4526(95)00789-X} {\bibfield  {journal} {\bibinfo  {journal}
  {Physica B}\ }\textbf {\bibinfo {volume} {219-220}},\
  \bibinfo {pages} {493} (\bibinfo {year} {1996})}\BibitemShut {NoStop}%
\bibitem [{\citenamefont {Siegle}\ \emph {et~al.}(1997)\citenamefont {Siegle},
  \citenamefont {Kaczmarczyk}, \citenamefont {Filippidis}, \citenamefont
  {Litvinchuk}, \citenamefont {Hoffmann},\ and\ \citenamefont
  {Thomsen}}]{Siegle1997}%
  \BibitemOpen
  \bibfield  {author} {\bibinfo {author} {\bibfnamefont {H.}~\bibnamefont
  {Siegle}}, \bibinfo {author} {\bibfnamefont {G.}~\bibnamefont {Kaczmarczyk}},
  \bibinfo {author} {\bibfnamefont {L.}~\bibnamefont {Filippidis}}, \bibinfo
  {author} {\bibfnamefont {A.~P.}\ \bibnamefont {Litvinchuk}}, \bibinfo
  {author} {\bibfnamefont {A.}~\bibnamefont {Hoffmann}}, \ and\ \bibinfo
  {author} {\bibfnamefont {C.}~\bibnamefont {Thomsen}},\ }\href {\doibase
  10.1103/PhysRevB.55.7000} {\bibfield  {journal} {\bibinfo  {journal} {Phys.
  Rev. B}\ }\textbf {\bibinfo {volume} {55}},\ \bibinfo {pages} {7000}
  (\bibinfo {year} {1997})}\BibitemShut {NoStop}%
\bibitem [{\citenamefont {Davydov}\ \emph {et~al.}(1998)\citenamefont
  {Davydov}, \citenamefont {Kitaev}, \citenamefont {Goncharuk}, \citenamefont
  {Smirnov}, \citenamefont {Graul}, \citenamefont {Semchinova}, \citenamefont
  {Uffmann}, \citenamefont {Smirnov}, \citenamefont {Mirgorodsky},\ and\
  \citenamefont {Evarestov}}]{Davydov1998}%
  \BibitemOpen
  \bibfield  {author} {\bibinfo {author} {\bibfnamefont {V.}~\bibnamefont
  {Davydov}}, \bibinfo {author} {\bibfnamefont {Y.}~\bibnamefont {Kitaev}},
  \bibinfo {author} {\bibfnamefont {I.}~\bibnamefont {Goncharuk}}, \bibinfo
  {author} {\bibfnamefont {A.}~\bibnamefont {Smirnov}}, \bibinfo {author}
  {\bibfnamefont {J.}~\bibnamefont {Graul}}, \bibinfo {author} {\bibfnamefont
  {O.}~\bibnamefont {Semchinova}}, \bibinfo {author} {\bibfnamefont
  {D.}~\bibnamefont {Uffmann}}, \bibinfo {author} {\bibfnamefont
  {M.}~\bibnamefont {Smirnov}}, \bibinfo {author} {\bibfnamefont
  {A.}~\bibnamefont {Mirgorodsky}}, \ and\ \bibinfo {author} {\bibfnamefont
  {R.}~\bibnamefont {Evarestov}},\ }\href {\doibase 10.1103/PhysRevB.58.12899}
  {\bibfield  {journal} {\bibinfo  {journal} {Phys. Rev. B}\ }\textbf {\bibinfo
  {volume} {58}},\ \bibinfo {pages} {12899} (\bibinfo {year}
  {1998})}\BibitemShut {NoStop}%
\bibitem [{\citenamefont {Liu}\ \emph {et~al.}(2007)\citenamefont {Liu},
  \citenamefont {Gu},\ and\ \citenamefont {Liu}}]{Liu2007}%
  \BibitemOpen
  \bibfield  {author} {\bibinfo {author} {\bibfnamefont {B.}~\bibnamefont
  {Liu}}, \bibinfo {author} {\bibfnamefont {M.}~\bibnamefont {Gu}}, \ and\
  \bibinfo {author} {\bibfnamefont {X.}~\bibnamefont {Liu}},\ }\href {\doibase
  10.1063/1.2800792} {\bibfield  {journal} {\bibinfo  {journal} {Appl. Phys.
  Lett.}\ }\textbf {\bibinfo {volume} {91}},\ \bibinfo {pages} {172102}
  (\bibinfo {year} {2007})}\BibitemShut {NoStop}%
\bibitem [{\citenamefont {Gillan}\ and\ \citenamefont
  {Jacobs}(1983)}]{Gillan1983}%
  \BibitemOpen
  \bibfield  {author} {\bibinfo {author} {\bibfnamefont {M.~J.}\ \bibnamefont
  {Gillan}}\ and\ \bibinfo {author} {\bibfnamefont {P.~W.~M.}\ \bibnamefont
  {Jacobs}},\ }\href {http://prb.aps.org/abstract/PRB/v28/i2/p759\_1}
  {\bibfield  {journal} {\bibinfo  {journal} {Phys. Rev. B}\ }\textbf {\bibinfo
  {volume} {28}} (\bibinfo {year} {1983})}\BibitemShut {NoStop}%
\bibitem [{\citenamefont {Walsh}\ \emph {et~al.}(2011)\citenamefont {Walsh},
  \citenamefont {Sokol},\ and\ \citenamefont {Catlow}}]{Walsh2011a}%
  \BibitemOpen
  \bibfield  {author} {\bibinfo {author} {\bibfnamefont {A.}~\bibnamefont
  {Walsh}}, \bibinfo {author} {\bibfnamefont {A.~A.}\ \bibnamefont {Sokol}}, \
  and\ \bibinfo {author} {\bibfnamefont {C.~R.~A.}\ \bibnamefont {Catlow}},\
  }\href {\doibase 10.1103/PhysRevB.83.224105} {\bibfield  {journal} {\bibinfo
  {journal} {Phys. Rev. B}\ }\textbf {\bibinfo {volume} {83}},\ \bibinfo
  {pages} {224105} (\bibinfo {year} {2011})}\BibitemShut {NoStop}%
\bibitem [{\citenamefont {Sanati}\ and\ \citenamefont
  {Estreicher}(2004)}]{Sanati2004}%
  \BibitemOpen
  \bibfield  {author} {\bibinfo {author} {\bibfnamefont {M.}~\bibnamefont
  {Sanati}}\ and\ \bibinfo {author} {\bibfnamefont {S.~K.}\ \bibnamefont
  {Estreicher}},\ }\href {\doibase 10.1088/0953-8984/16/28/L02} {\bibfield
  {journal} {\bibinfo  {journal} {J. Phys.: Condens. Matter}\ }\textbf
  {\bibinfo {volume} {16}},\ \bibinfo {pages} {L327} (\bibinfo {year}
  {2004})}\BibitemShut {NoStop}%
\bibitem [{\citenamefont {Peshek}\ \emph {et~al.}(2008)\citenamefont {Peshek},
  \citenamefont {Angus},\ and\ \citenamefont {Kash}}]{Peshek2008}%
  \BibitemOpen
  \bibfield  {author} {\bibinfo {author} {\bibfnamefont {T.~J.}\ \bibnamefont
  {Peshek}}, \bibinfo {author} {\bibfnamefont {J.~C.}\ \bibnamefont {Angus}}, \
  and\ \bibinfo {author} {\bibfnamefont {K.}~\bibnamefont {Kash}},\ }\href
  {\doibase 10.1016/j.jcrysgro.2008.09.203} {\bibfield  {journal} {\bibinfo
  {journal} {J. Cryst. Growth}\ }\textbf {\bibinfo {volume} {311}},\ \bibinfo
  {pages} {185} (\bibinfo {year} {2008})}\BibitemShut {NoStop}%
\bibitem [{\citenamefont {Jacob}\ and\ \citenamefont
  {Rajitha}(2009)}]{Jacob2009}%
  \BibitemOpen
  \bibfield  {author} {\bibinfo {author} {\bibfnamefont {K.}~\bibnamefont
  {Jacob}}\ and\ \bibinfo {author} {\bibfnamefont {G.}~\bibnamefont
  {Rajitha}},\ }\href {\doibase 10.1016/j.jcrysgro.2009.05.016} {\bibfield
  {journal} {\bibinfo  {journal} {J. Cryst. Growth}\ }\textbf {\bibinfo
  {volume} {311}},\ \bibinfo {pages} {3806} (\bibinfo {year}
  {2009})}\BibitemShut {NoStop}%
\bibitem [{\citenamefont {Momma}\ and\ \citenamefont
  {Izumi}(2011)}]{Momma2011}%
  \BibitemOpen
  \bibfield  {author} {\bibinfo {author} {\bibfnamefont {K.}~\bibnamefont
  {Momma}}\ and\ \bibinfo {author} {\bibfnamefont {F.}~\bibnamefont {Izumi}},\
  }\href {\doibase 10.1107/S0021889811038970} {\bibfield  {journal} {\bibinfo
  {journal} {J. Appl. Crystallogr.}\ }\textbf {\bibinfo {volume} {44}},\
  \bibinfo {pages} {1272} (\bibinfo {year} {2011})}\BibitemShut {NoStop}%

    
\end{thebibliography}

%

\end{document}